\documentclass[a4paper,fleqn]{cas-dc}

\usepackage[authoryear]{natbib}

\def\tsc#1{\csdef{#1}{\textsc{\lowercase{#1}}\xspace}}
\tsc{WGM}
\tsc{QE}

\usepackage{nicefrac}
\newcommand{\Sabs}[1]{\left\lceil #1 \right\rfloor}

\newcommand{\abs}[1]{\left| #1 \right|}

\PassOptionsToPackage{switch}{lineno}

\begin{document}
\let\WriteBookmarks\relax
\def\floatpagepagefraction{1}
\def\textpagefraction{.001}

\shorttitle{}    

\shortauthors{}  

\title [mode = title]{A Control Framework for Induced Seismicity Mitigation: Application in the Groningen Gas Field, the Netherlands}


\author[1]{Diego Guti\'errez-Oribio}[orcid=https://orcid.org/0000-0002-5895-3416]
\ead{diego.gutierrez@ensta.fr}
\credit{Conceptualization, Formal Analysis, Methodology, Software, Validation, Visualization, Writing – Original Draft Preparation}
\affiliation[1]{organization={IMSIA (UMR 9219), CNRS, EDF, ENSTA Paris, Institut Polytechnique de Paris},
            addressline={828, Boulevard des Maréchaux}, 
            postcode={91762}, 
            state={Palaiseau},
            country={France}}

\author[1]{Ioannis Stefanou}[orcid=https://orcid.org/0000-0002-4552-7717]
\cormark[1]
\ead{ioannis.stefanou@ensta.fr}
\credit{Conceptualization, Formal Analysis, Funding Acquisition, Methodology, Project Administration, Software, Writing – Review \& Editing}
\cortext[1]{Corresponding author}


\begin{abstract}
Induced seismicity associated with gas production poses major operational and societal challenges, as illustrated by the Groningen gas field in the Netherlands. This work addresses the problem of regulating seismic activity rather than only forecasting it under prescribed production scenarios. A depth-averaged pore-pressure diffusion model is coupled with seismicity-rate (SR) dynamics and a non-homogeneous Poisson process that generates synthetic earthquake catalogs with event times, locations, and magnitudes. Regional SR is reconstructed from finite-window event counts and used by a robust super-twisting feedback controller to compute well-rate commands to regulate the SR toward a desired reference while satisfying operational requirements, including prescribed production constraints. The proposed control architecture explicitly accounts for injection and extraction flux limits (actuator saturation), ensuring that the applied well rates respect the modeled constraints. The well fluxes generated by the controller are updated at discrete-time intervals (digital control). Therefore, the main contribution is the integration of stochastic seismic catalog generation, and constrained sampled-data feedback to mitigate induced seismicity while accounting for production requirements. Calibration agreement is quantified against Groningen observations, and extraction-only feedback is evaluated against the historical no-control baseline using 50 stochastic realizations for each case. The results show substantial SR reduction in an extraction-only configuration, with an explicit production--seismicity compromise, and stronger mitigation when independent compensating-injection channels are available. The latter scenario is presented as a conceptual control-allocation study rather than a field-ready prescription.
\end{abstract}

\begin{graphicalabstract}
  \centering
  \includegraphics[width=14.0cm,keepaspectratio]{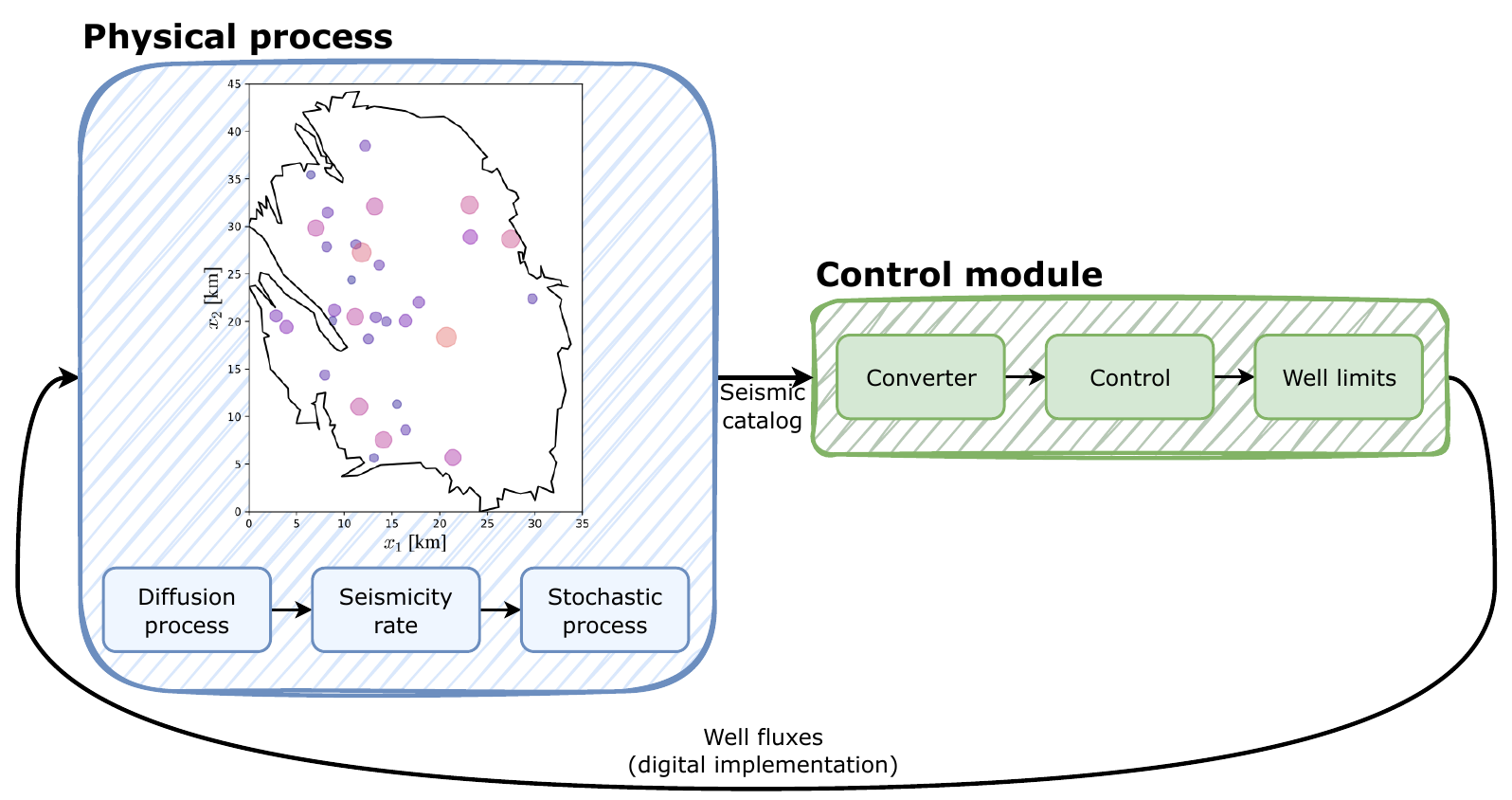}
\end{graphicalabstract}

\begin{highlights}
  \item Feedback control mitigates seismicity under production constraints.
  \item A diffusion--seismicity cascade generates stochastic earthquake catalogs.
  \item Catalog-based seismicity estimates provide feedback for rate regulation.
  \item Demand, well-rate bounds, and sampled-data actuation are enforced.
  \item Sampling, gain tuning, and compensating injection are numerically assessed.
\end{highlights}


\begin{keywords}
Induced seismicity mitigation \sep Gas production and control \sep Robust feedback control
\end{keywords}

\maketitle

\section{Introduction}
\label{sec:Intro}

Induced seismicity is a major operational and societal concern because subsurface activities perturb pore pressure and stress, potentially reactivating faults and triggering earthquakes. It can create safety risks and structural damage, constrain operations through regulatory protocols and rate limitations, and ultimately compromise the license to operate.

In North America, the marked increase in seismicity in the U.S. mid-continent since approximately 2009 has been linked to high-volume wastewater disposal associated with hydrocarbon production \citep{b:Ellsworth2013,b:Keranen-Savage-Abers-Cochran-2013,b:KeranenEtAl2014,b:WeingartenEtAl2015}. Related concerns have arisen in western Canada, the United Kingdom, and China, where hydraulic fracturing and associated fluid disposal have been connected to felt and, in some cases, damaging seismicity \citep{b:AtkinsonEtAl2016,b:SchultzEtAl2020}. In Europe, the Groningen gas field in the Netherlands is a well-documented example of production-induced seismicity associated with long-term reservoir depletion, with substantial consequences for the built environment and public acceptance of gas production \citep{b:VanThienenVisserBreunese2015}.

\begin{figure}[pos=htbp]
  \centering
  \includegraphics[width=6.0cm,keepaspectratio]{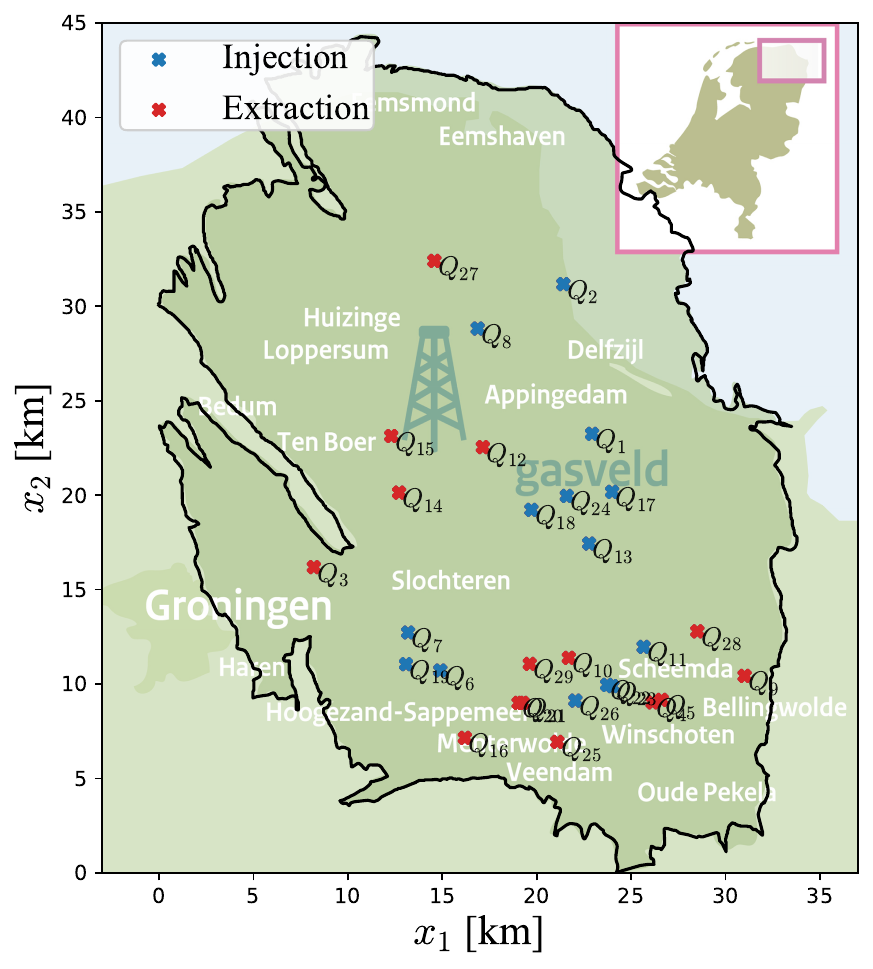}
  \caption{Location of the Groningen gas field in the northeastern Netherlands and wells locations. The blue/red assignment is used only for the conceptual production--injection study of Scenario~2 and does not indicate an existing field configuration. Background map: \url{https://zoek.officielebekendmakingen.nl/stcrt-2017-28922.html}.}
  \label{fig:Groningen}
\end{figure}

The Groningen gas field, located in the northeastern Netherlands, is the largest onshore natural-gas field in Europe and one of the largest worldwide, with an estimated 2{,}900 billion m$^3$ of recoverable gas. Production-induced earthquakes were first recorded in 1991 and subsequently caused structural damage and serious public concern. In June 2023, the Dutch government announced that extraction would cease by October 1, 2023, leaving approximately 470 billion m$^3$ of gas in the field \citep{NAM2016GPM,b:Groningen1,b:Groningen2}. Figure~\ref{fig:Groningen} locates the field and shows the well locations.

Groningen has been studied extensively because of its societal importance and the availability of high-quality monitoring and production data \citep{b:SMITH2022117697,b:doi.org/10.1029/2023GL105455,b:10.1785/0220230179,b:https://doi.org/10.1029/2024GL110139,b:https://doi.org/10.1029/2020JB020013}. Most Groningen-specific studies address the forward problem of forecasting seismicity under prescribed production histories, interpreting its geomechanical origin, or optimizing a prescribed production distribution. The present objective is different: to close the loop from observed seismic events to constrained well-rate commands that regulate seismic activity while accounting for production requirements.

At the fault scale, feedback control has been investigated to stabilize frictionally unstable systems and promote controlled slow slip instead of rapid dynamic failure \citep{b:Stefanou2019,b:https://doi.org/10.1029/2021JB023410,b:Gutierrez-Tzortzopoulos-Stefanou-Plestan-2022,b:Gutierrez-Orlov-Plestan-Stefanou-VSS2022,b:Gutierrez-Orlov-Stefanou-Plestan-2023,b:Gutierrez-Stefanou-Plestan-2024}. At the reservoir scale, robust control and optimization methods have been developed to regulate seismicity-rate (SR) models coupled with fluid diffusion \citep{b:https://doi.org/10.1002/nag.3923,b:Gutierrez-Stefanou-2024,b:Gutierrez-Stefanou-2025,b:KIM2025103396}. These studies establish the control-theoretic basis, but generally assume a deterministic SR output and idealized continuous-time actuation.

The present work contributes an integrated catalog-driven control framework. A pressure-diffusion--SR cascade is coupled with a stochastic event-generation layer that produces event times, locations, and magnitudes. A regional SR measurement is reconstructed from the resulting catalog and used by a robust feedback controller. The commanded well rates satisfy componentwise bounds, may enforce a prescribed production demand, and are implemented through a zero-order hold. The closed loop is evaluated against the historical no-control baseline and over ensembles of stochastic catalog realizations. Thus, the contribution is not the isolated use of the diffusion--SR coupled system, stochastic catalogs, or sliding-mode control, but their integration into a constrained sampled-data feedback architecture for the Groningen case study.

The paper is organized as follows. Section~\ref{sec:Methodology} presents the geological setting, physical model, stochastic catalog generator, control design, and numerical implementation. Section~\ref{sec:Results} presents the numerical results, Section~\ref{sec:Discussion} discusses their interpretation, limitations, and operational implications, and Section~\ref{sec:Conclusions} summarizes the conclusions.

\section{Methodology}
\label{sec:Methodology}

\begin{figure*}[pos=tbp]
  \centering
  \includegraphics[width=14.0cm,keepaspectratio]{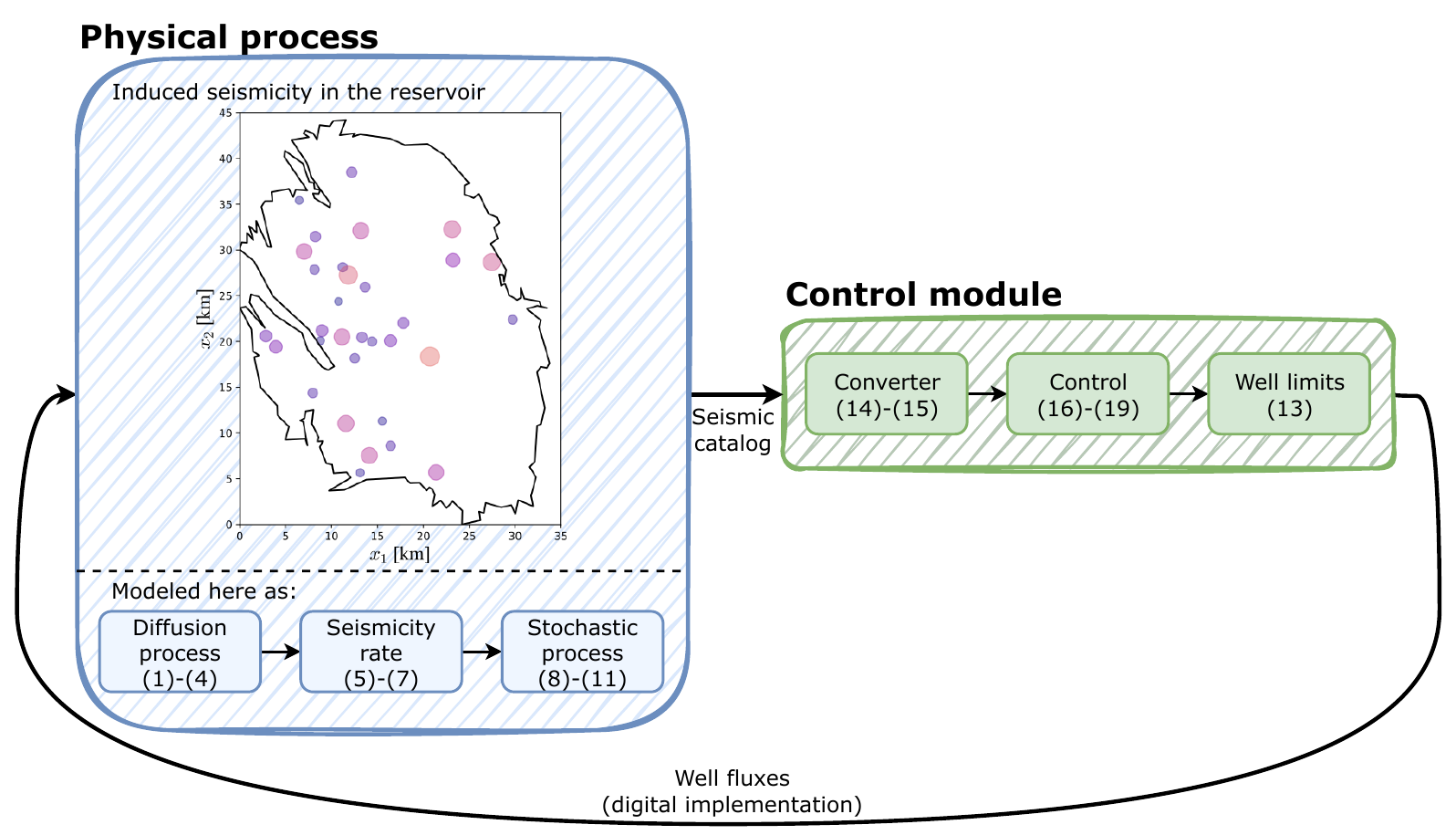}
  \caption{Closed-loop methodology. The physical model maps well rates to a continuous SR field and then to a stochastic seismic catalog. At each control update, the catalog is converted into a regional SR estimate, the feedback command is computed, well-rate limits are imposed, and the resulting command is held constant until the next update.}
  \label{fig:block}
\end{figure*}

Figure~\ref{fig:block} summarizes the closed-loop methodology. The physical process combines pore-pressure diffusion, SR dynamics, and stochastic event generation. The control module converts the seismic catalog into a filtered regional SR estimate and computes bounded well-rate commands at discrete update times.

\subsection{Regional geological setting}

The Groningen field is located within the Southern Permian Basin. Gas is hosted mainly in Rotliegend reservoir sandstones and sealed by the overlying Zechstein evaporitic sequence. The reservoir is intersected by faults, and long-term pressure depletion produces reservoir compaction and stress changes that can promote fault reactivation. These features are treated in dedicated Groningen geological and geomechanical studies \citep{DeJagerVisser2017,VanWeesEtAl2017}.

\subsection{Physical process}

\subsubsection{Model setup and calibration}

Fluid injection and extraction generate pressure gradients that drive flow within the reservoir and deform the surrounding porous rock. This hydro-mechanical response is classically described by Biot's theory \citep{b:Biot-1941}. When the operations vary sufficiently slowly for inertial effects to be neglected and the volumetric strain contribution is omitted, the pore-pressure perturbation is approximated by the diffusion equation \citep{b:Cheng2016}
\begin{equation}
  u_t=-\frac{1}{\beta}\nabla\!\cdot q+\frac{1}{\beta}s,
  \label{eq:diff}
\end{equation}
where $u=u(x,t)$ is the pore-pressure change, $x\in V$ is the spatial coordinate, $t\geq0$ is time, $s=s(x,t)$ is the volumetric source density, and $\beta>0$ is the effective compressibility of the rock--fluid mixture. Darcy's law gives
\begin{equation}
  q=-\frac{k}{\eta}\nabla u,
  \label{eq:flux}
\end{equation}
where $q=q(x,t)$ is the Darcy flux, $k>0$ is permeability, and $\eta>0$ is dynamic viscosity.

All simulations use the homogeneous no-flow boundary condition $q(x,t)\cdot n(x)=0$, $x\in\partial V$, where $n(x)$ is the outward unit normal (see \cite{b:SMITH2022117697,b:doi.org/10.1029/2023GL105455,b:10.1785/0220230179,b:https://doi.org/10.1029/2024GL110139,b:https://doi.org/10.1029/2020JB020013} for further discussion of boundary conditions in Groningen). The $n=29$ production locations are represented by distributed sources over subvolumes $V_i^*\subset V$,
\begin{equation}
  s(x,t)=\mathcal{B}(x)Q(t),
  \label{eq:input}
\end{equation}
where $Q=[Q_1,\ldots,Q_n]^T\in\mathbb{R}^n$ is the well-rate vector and $\mathcal{B}=[\mathcal{B}_1,\ldots,\mathcal{B}_n]\in\mathbb{R}^{1\times n}$, with
\begin{equation}
  \mathcal{B}_i(x)=
  \begin{cases}
    1/V_i^*, & x\in V_i^*,\\
    0, & x\notin V_i^*,
  \end{cases}
  \qquad i=1,\ldots,n.
  \label{eq:B}
\end{equation}
Positive and negative values of $Q_i$ denote injection and extraction, respectively. 

It is well established that fluid injection and extraction can reactivate pre-existing faults, or bring fractures closer to failure, thereby generating induced earthquakes. This seismicity arises from stress changes associated with pore-pressure diffusion and fluid migration, which can promote fault slip and increase seismic activity (see \cite{b:Rubinstein-Mahani-2015,b:Zastrow-2019,b:Keranen-Savage-Abers-Cochran-2013} in general, and \cite{b:SMITH2022117697,b:doi.org/10.1029/2023GL105455,b:10.1785/0220230179} for Groningen in particular). In this sense, injection/production operations alter the seismicity rate (SR), i.e., the number of seismic events expected to occur within a given time window and region.

Following \cite{Dieterich1994,Segall2015}, the normalized SR, $R^n=R^n(x,t)$, satisfies
\begin{equation}
  R^n_t=\frac{R^n}{t_a}\left(\frac{\dot{\tau}}{\dot{\tau}_0}-R^n\right),
  \label{eq:SR1}
\end{equation}
where $\dot{\tau}=\dot{\tau}(x,t)$ is the Coulomb stressing rate, $\dot{\tau}_0>0$ is its background value, and $t_a>0$ is a characteristic decay time. The simplified stressing rate
\begin{equation}
  \dot{\tau}(x,t)=\dot{\tau}_0-fu_t(x,t)
  \label{eq:dCFS}
\end{equation}
is adopted, where $f>0$ is an effective friction coefficient. Equation~\eqref{eq:dCFS} is a simplified closure rather than a complete geomechanical model of Groningen. It neglects reservoir compaction, poroelastic stress redistribution, fault geometry and frictional heterogeneity, spatially varying stress paths, and possible nonlinear or history-dependent production--stress coupling. These mechanisms are relevant in Groningen and are represented with greater physical detail in geological, geomechanical, and stress-based studies \citep{VanWeesEtAl2017,b:SMITH2022117697,b:doi.org/10.1029/2023GL105455,b:https://doi.org/10.1029/2024GL110139,b:https://doi.org/10.1029/2020JB020013}. The simplification is retained to obtain a tractable pressure--seismicity cascade for evaluating the closed-loop architecture. A higher-fidelity model may replace this block without changing the catalog-estimation and control layers. Nevertheless, no claim is made that the robustness of the presented controller compensates for arbitrary missing physics. The robustness statement is restricted to the admissible bounded perturbation class specified by the control model.

Defining the unnormalized SR density as $R(x,t)=R^n(x,t)R^*(x)$, with $R^*(x)>0$ as the background SR density, gives
\begin{equation}
  R_t=R\left[-\gamma_1u_t-\gamma_2\left(\frac{R}{R^*}-1\right)\right],
  \label{eq:SR}
\end{equation}
with $\gamma_1=\frac{f}{t_a\dot{\tau}_0}$, $\gamma_2=\frac{1}{t_a}$. The parameter $\gamma_1$ has units of inverse pressure and quantifies the pressure-rate sensitivity, whereas $\gamma_2$ has units of inverse time and sets the relaxation rate toward $R^*$. With the sign convention in \eqref{eq:dCFS}, extraction-induced pressure decrease, $u_t<0$, increases the SR. Table~\ref{tab:notation} summarizes the principal quantities of both the diffusion and SR systems.

\begin{table*}[pos=tbp]
\centering
\caption{Principal physical, stochastic, and control quantities.}
\label{tab:notation}
\small
\begin{tabular}{lll}
\hline
Symbol & Definition & Units \\
\hline
$u$ & Pore-pressure change & pressure \\
$q$ & Darcy flux & length/time \\
$s$ & Distributed volumetric source density & time$^{-1}$ \\
$\beta$, $k$, $\eta$ & Effective compressibility, permeability, dynamic viscosity & pressure$^{-1}$, length$^2$, pressure$\cdot$time \\
$Q_i$ & Volumetric flow rate at well $i$ & volume/time \\
$R$, $R^*$ & SR density and background SR density & events/(time$\cdot$volume) \\
$\gamma_1$, $\gamma_2$ & Pressure-rate sensitivity and SR relaxation rate & pressure$^{-1}$, time$^{-1}$ \\
$y_R$, $r_R$ & Estimated regional SR and its reference & events/(time$\cdot$volume) \\
$\Delta t_c$, $\tau$ & Control update period and filter time constant & time \\
$\sigma_R$, $\nu$ & Tracking error and controller internal state & pressure, pressure/time \\
$B_0$ & Nominal control matrix & pressure/volume \\
$l$ & Controller parameter & -- \\
$k_1,k_2$ & Proportional and integral controller gains & pressure$^{-l/(1-l)}$ $\cdot$ time$^{-1}$, pressure$^{-2l/(1-l)}$ $\cdot$ time$^{-2}$ \\
$k_3,\rho$ & Leakage gain and saturation-dependent coefficient & pressure$^{(1+l)/(1-l)}$, pressure$^{-(1+l)/(1-l)}$ $\cdot$ time$^{-1}$\\
\hline
\end{tabular}
\end{table*}

Following \cite{Segall2015,Dieterich1994,b:SMITH2022117697,b:doi.org/10.1029/2023GL105455,b:10.1785/0220230179,b:https://doi.org/10.1029/2024GL110139,b:https://doi.org/10.1029/2020JB020013,b:kim2023,https://doi.org/10.1029/2019JB019134,https://doi.org/10.1029/2018WR023587}, system \eqref{eq:SR} is a well-established model for describing the number of seismic events per unit time in a given region induced by fluid injection/extraction. This model has been applied to several reservoirs, including Groningen, Otaniemi, Pohang, Quest, and Oklahoma, among others \citep{b:SMITH2022117697,b:doi.org/10.1029/2023GL105455,b:10.1785/0220230179,b:https://doi.org/10.1029/2024GL110139,b:https://doi.org/10.1029/2020JB020013,b:kim2023,https://doi.org/10.1029/2019JB019134,b:AcostaEtAl2025Flow2Quake,https://doi.org/10.1029/2018WR023587}. 

Furthermore, \cite{b:https://doi.org/10.1029/2024JB030243} compared the SR model with numerical simulations of induced earthquakes in a discrete fault network governed by rate-and-state friction. They show that, despite its simplifying assumptions, the SR model can reproduce the \emph{overall} (qualitative) spatio-temporal evolution of seismicity produced by the discrete-fault system over a wide range of conditions, while emphasizing that finite-size effects and fault-network limitations can bias the interpretation of the inferred SR parameters.

\begin{figure}[pos=htbp]
  \centering
  \includegraphics[width=8.0cm,keepaspectratio]{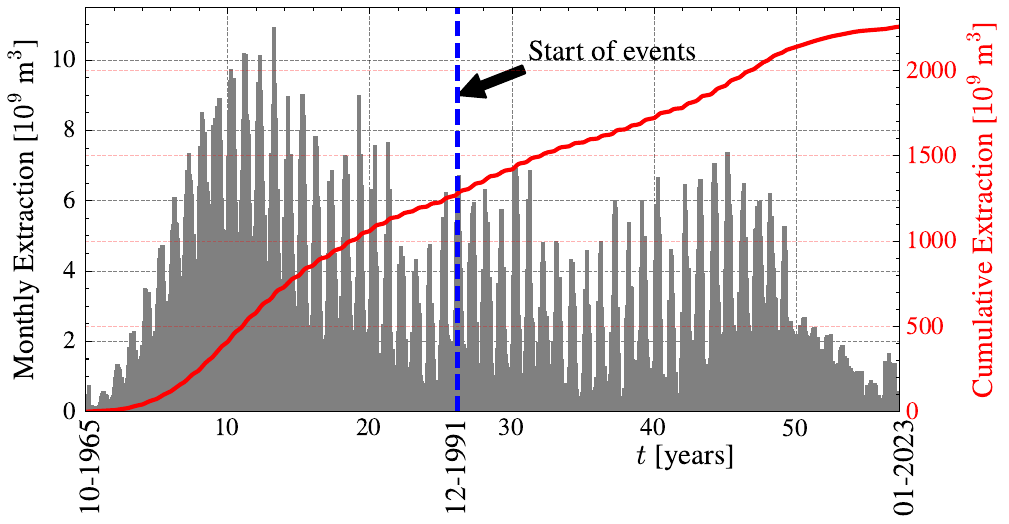}
  \caption{Historical Groningen gas extraction used as the no-control input and as the production reference in the controlled scenarios. The bars show monthly extraction, the solid curve shows cumulative extraction, and the vertical markers identify the beginning of the seismic catalog and the end of the simulation period.}
  \label{fig:extraction}
\end{figure}

\begin{figure*}[pos=tbp]
  \centering
  \includegraphics[width=\textwidth,keepaspectratio]{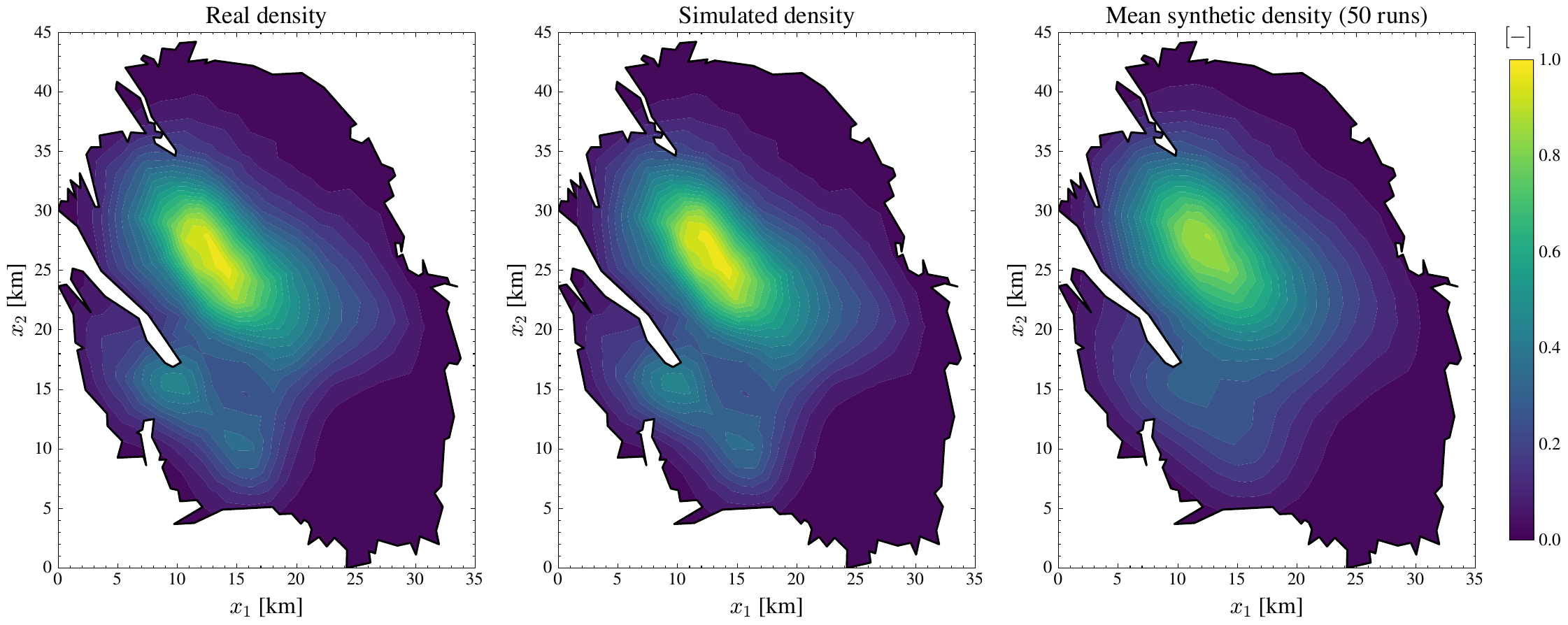}
  \caption{Spatial comparison of the observed density, the calibrated deterministic density, and the mean synthetic density obtained from 50 stochastic catalogs over December 1991--January 2023. All three densities are normalized under the same reference. Spatial variability across 50 no-control realizations is quantified in Table~\ref{tab:ensemble}.}
  \label{fig:density}
\end{figure*}

\begin{figure}[pos=htbp]
  \centering
  \hspace*{-0.2cm}\includegraphics[width=8.0cm,keepaspectratio]{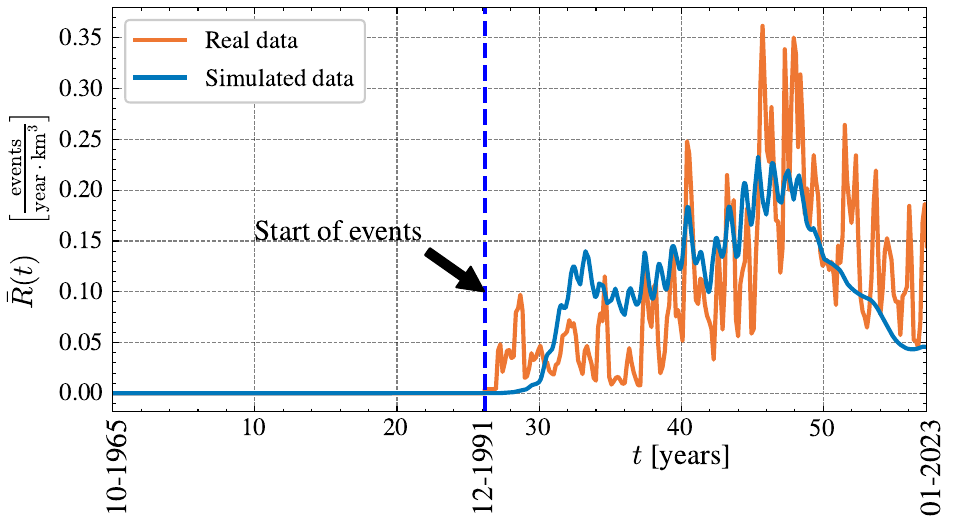}
  \includegraphics[width=8.0cm,keepaspectratio]{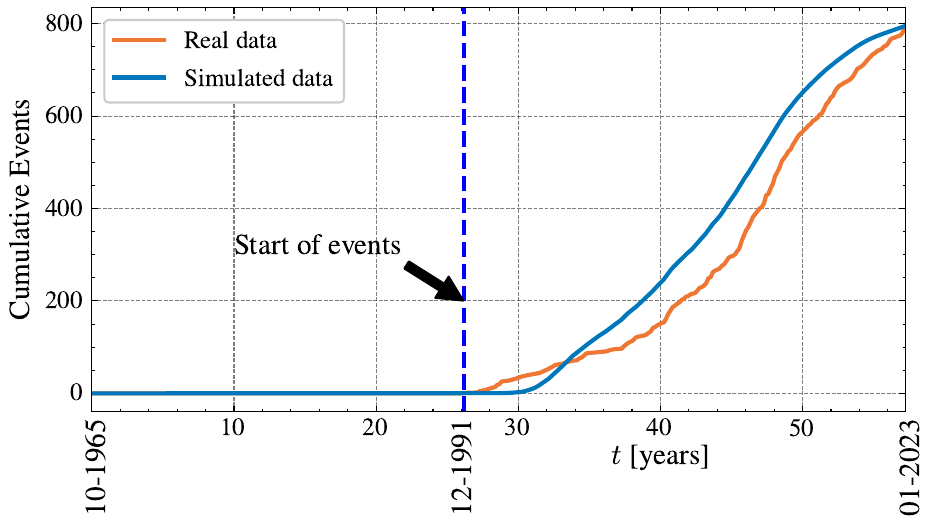}
  \caption{Temporal calibration agreement of the deterministic model. The upper panel compares the observed and simulated reservoir-average SR, the lower panel compares their cumulative event counts from the beginning of the catalog.}
  \label{fig:validation}
\end{figure}

In the absence of injection or extraction, $u_t=0$ and $R\to R^*$. Under historical Groningen extraction, $u_t<0$ and the model predicts an increase in SR. Figure~\ref{fig:extraction} shows the monthly and cumulative extraction history from October 1965 to January 2023. The catalog contains 790 events from December 1991 to January 2023 above the selected completeness magnitude $M_c=1.0$. Their spatial distribution and the corresponding reservoir-average SR are shown in Figs.~\ref{fig:density} and \ref{fig:validation}.

The hydraulic parameters are selected from Groningen studies \citep{b:SMITH2022117697,b:doi.org/10.1029/2023GL105455,b:10.1785/0220230179,b:https://doi.org/10.1029/2024GL110139,b:https://doi.org/10.1029/2020JB020013}. The effective hydraulic diffusivity and effective compressibility are $c_{hy}=k/(\beta\eta)=4.48\times10^{-2}$~km$^2$/hr and $\beta=4.2\times10^{-1}$~MPa$^{-1}$, respectively. The SR parameters are calibrated from the observed spatial seismicity density, giving $\gamma_1(x)=4.0$~MPa$^{-1}$, $\gamma_2=4.6\times10^{-8}$~hr$^{-1}$, and $R^*(x)=4.11\times10^{-5}d(x)$~events/(year$\cdot$km$^3$), where $d(x)$ is the normalized observed density.

Quantitative agreement is assessed through the normalized root-mean-square error of the cumulative event curve,
\begin{equation*}
  \mathrm{NRMSE}_C=
  \frac{\left[\frac{1}{N_t}\sum_{j=1}^{N_t}
  \left(C_{\mathrm{sim}}(t_j)-C_{\mathrm{obs}}(t_j)\right)^2\right]^{1/2}}
  {\max_j C_{\mathrm{obs}}(t_j)-\min_j C_{\mathrm{obs}}(t_j)}.
  \label{eq:nrmse}
\end{equation*}
The calibrated model gives $\mathrm{NRMSE}_C=0.09$ and $~795$ final cumulative events, resulting in a final-count relative error of $0.71\%$. The Pearson correlation between the observed and simulated normalized density maps is $0.9999$. Since the observed density enters the parameter calibration, this correlation demonstrates spatial consistency rather than independent predictive validation. Accordingly, the model is used as a reduced-order control testbed that reproduces the principal temporal and spatial trends. It is not claimed to be a high-fidelity Groningen forecasting model. 

All the presented results in this paper were performed over a depth-averaged 2D model using finite-elements method (see \cite{b:Gutierrez-Stefanou-2025} for details on the numerical implementation of the model). Therefore, the model does not resolve vertical layering, fault offsets, or local stress concentrations.

\subsubsection{Synthetic catalog generation}
\label{sec:synthetic}

To generate a synthetic catalog of discrete seismic events (time, location, and magnitude) from the continuous seismicity-rate field $R(x,t)$ of the cascade system \eqref{eq:diff}--\eqref{eq:SR} over $t \in [t_1,t_2]$, event occurrence is modeled as a non-homogeneous Poisson point process \citep{b:Ross2013Simulation} sampled over the space--time domain $V \times [t_1,t_2]$. The process is fully specified by the intensity (seismicity-rate density) $R(x,t)$, which is interpreted as the expected number of events per unit volume and unit time at location $x\in V$ and time $t$.

A central quantity is the total event rate
\begin{equation}
\Lambda(t) \;=\; \int_{V} R(x,t)\, \mathrm{d}x,
\label{eq:event_rate}
\end{equation}
which represents the expected number of events per unit time over the entire domain. The expected number of events in the interval $[t_1,t_2]$ is
\begin{equation}
\Delta\Lambda \;=\; \int_{t_1}^{t_2} \Lambda(t)\,\mathrm{d}t,
\label{eq:cum_intensity}
\end{equation}
from which the realized number of events is sampled as $N \sim \mathrm{Poisson}(\Delta\Lambda)$. For each time interval $[t_1,t_2]$, $\Lambda(t)$ is approximated by a linear interpolation between its endpoint values $\Lambda(t_1)$ and $\Lambda(t_2)$. This is a reasonable approximation for sufficiently small time intervals and allows the $N$ event times $s_i$, $i=1,2,\ldots,N$, to be drawn in the interval $[t_1,t_2]$ by inversion of the cumulative intensity \eqref{eq:cum_intensity}. The drawn events are then sorted (see the Inverse Transform Method in \cite{b:Ross2013Simulation}).

Conditioned on an event time $s_i$, the event location is drawn according to the instantaneous spatial distribution of the rate field. The corresponding conditional density is
\begin{equation}
p(x \,|\, s_i) \;=\; \frac{R(x,s_i)}{\Lambda(s_i)},
\label{eq:conditional_density}
\end{equation}
so regions with larger values of $R(\cdot,s_i)$ are proportionally more likely to generate events. Sampling from this distribution is performed using an acceptance--rejection (``thinning'' \citep{b:Ross2013Simulation}) procedure: candidate points are proposed uniformly over $V$, and a candidate $x$ is accepted with probability $R(x,s_i)/R_{\max}(s_i)$, where $R_{\max}(s_i)$ is an upper bound on the intensity at time $s_i$. Repeating this procedure until acceptance produces samples distributed according to $p(x \,|\, s_i)$ without explicitly normalizing the field. In this way, the continuous spatial SR map at time $s_i$ is converted into discrete event locations while preserving the prescribed relative occurrence likelihood across the domain.

Finally, to obtain a complete synthetic catalog, a magnitude $M_i$ is independently assigned to each accepted event using a truncated Gutenberg--Richter (G--R) distribution \citep{b:GutenbergRichter1944Frequency,b:Aki1965MLEbvalue,b:li2026distinguishing},
\begin{equation}
\mathbb{P}(M \ge m) \;\propto\; 10^{-b m}, \qquad m\in[M_c,M_{\max}],
\label{eq:GR}
\end{equation}
sampled via the Inverse Transform Method. This adds a standard magnitude--frequency structure to the spatio-temporal occurrence model, enabling direct comparison with observed seismic catalogs. Notice that a tapered G--R distribution could better describe the magnitude statistics in Groningen, including potential stress dependencies of its parameters \citep{b:li2026distinguishing,b:https://doi.org/10.1029/2020JB020013}. However, the simpler distribution is retained because the present control module uses only event occurrences and not their magnitudes.

The resulting output is a set of events $\{(s_i, x_i, M_i)\}_{i=1}^{N}$ forming a statistically consistent realization of the prescribed intensity field $R(x,t)$. In particular, the synthetic catalog preserves both the time-varying domain-wide activity level through $\Lambda(t)$ and the instantaneous spatial distribution encoded by $R(\cdot,t)$, while naturally incorporating the intrinsic randomness of non-homogeneous Poisson event occurrence. 

\begin{figure*}[pos=tbp]
  \centering
  \includegraphics[width=\textwidth,keepaspectratio]{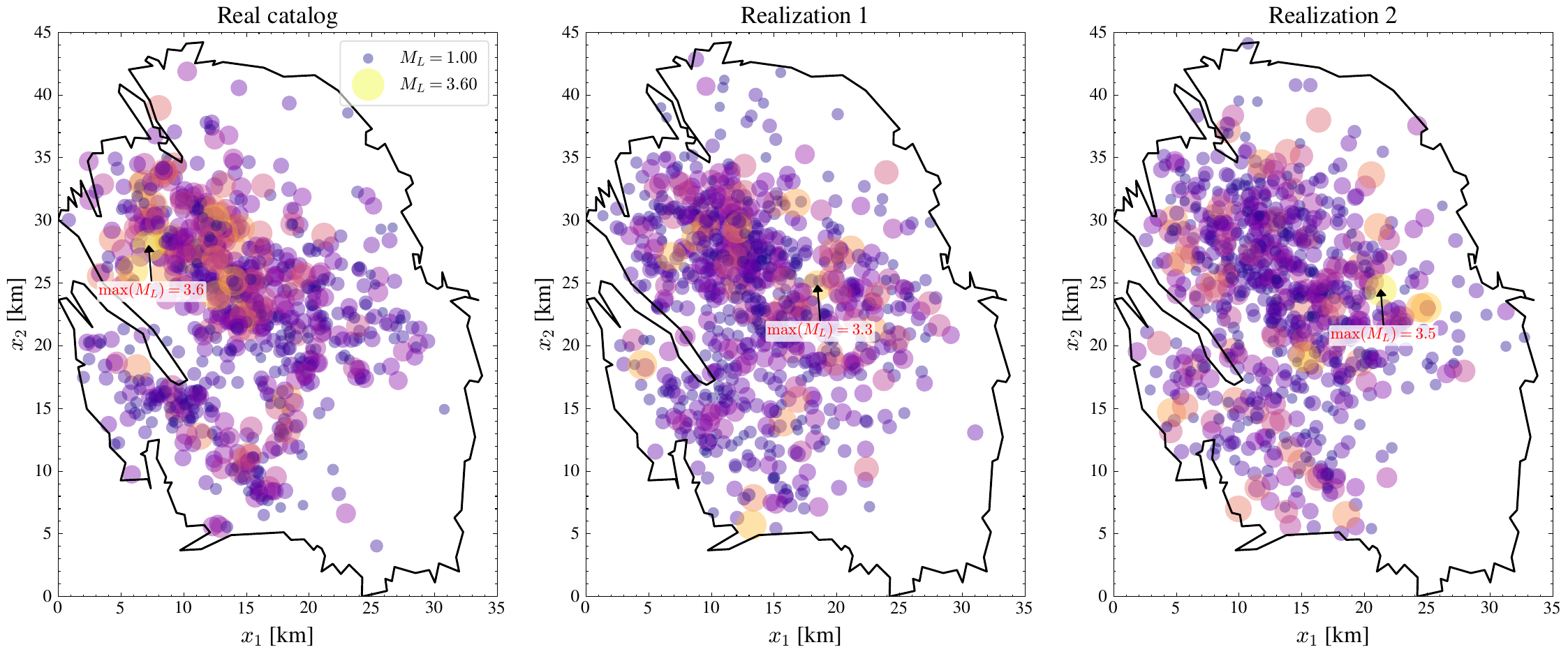}
  \caption{Observed Groningen catalog and two stochastic realizations generated from the calibrated deterministic SR field. Marker size and color represent local magnitude. The realizations illustrate catalog-to-catalog variability. Ensemble statistics are reported separately in Table~\ref{tab:ensemble}.}
  \label{fig:catalog}
\end{figure*}

The G--R parameters were estimated from the real Groningen seismic catalog, yielding $a=4.075$, $b=0.968$, $M_c=1.0$, and $M_{\max}=3.6$. These values were then used to drive the stochastic process. As an example, two synthetic catalogs (realizations) are shown in Fig.~\ref{fig:catalog}. Re-estimating the G--R parameters from these two catalogs yields $4.035$ and $4.121$ for $a$, and $1.028$ and $1.061$ for $b$, respectively, indicating that the synthetic magnitudes reproduce the overall magnitude--frequency trend of the original dataset well. Furthermore, the resulting seismic density is comparable to both the real and simulated densities, as shown in Fig.~\ref{fig:density} (right). It should be emphasized that, since event generation is stochastic, any single synthetic catalog represents one random realization of the underlying intensity field. Consequently, as the number of realizations increases, ensemble statistics (e.g., the mean cumulative number of events) converge to the reference solution provided by the deterministic model \eqref{eq:diff}--\eqref{eq:SR}. This consistency is quantified over 50 independent realizations in Table~\ref{tab:ensemble} and Fig.~\ref{fig:runs} (No control).

\subsection{Control module}

The total well-rate input is decomposed as
\begin{equation}
  Q(t)=Q_c(t)+Q_s(t),
  \label{eq:control+static}
\end{equation}
where $Q_s$ is a prescribed production component and $Q_c$ is the feedback component. Each rate is subject to the componentwise bounds $Q_{m_i}\leq Q_i\leq Q_{M_i}$,
\begin{equation}
  \operatorname{sat}(Q_i)=
  \begin{cases}
    Q_i, & Q_{m_i}<Q_i<Q_{M_i},\\
    Q_{M_i}, & Q_i\geq Q_{M_i},\\
    Q_{m_i}, & Q_i\leq Q_{m_i},
  \end{cases}
  \label{eq:sat}
\end{equation}
for $i=1,\ldots,n$. The objective is to regulate the reservoir-average SR from December 1991 onward while preserving the prescribed production requirement as far as the available actuation permits.

\subsubsection{Converter: discrete events to SR}

At each sampling instant $t_k$, let $n(t_k)$ denote the number of events observed in $V$ during $[t_{k-1},t_k)$. The raw reservoir-average SR estimate is
\begin{equation}
  \hat y_R(t_k)=\frac{n(t_k)}{|V|\Delta t_c},
  \label{eq:raw}
\end{equation}
where $|V|$ is the reservoir volume and $\Delta t_c=t_k-t_{k-1}>0$ is the control update period. Because finite-window event counts are noisy, an exponential moving average is used,
\begin{equation}
  y_R(t_k)=(1-\alpha)y_R(t_{k-1})+\alpha\hat y_R(t_k),
  \qquad
  \alpha=1-e^{-\Delta t_c/\tau},
  \label{eq:ema_disc}
\end{equation}
where $\tau>0$ is the filter time constant. Smaller $\tau$ gives a faster but noisier estimate, whereas larger $\tau$ increases smoothing and delay. The reference $r_R(t)>0$ specifies the desired regional SR level.

\subsubsection{Control design}

The tracking error is
\begin{equation}
  \sigma_R(t)=\frac{y_R(t)-r_R(t)}{\gamma_{1_0}R^*_0},
  \label{eq:error}
\end{equation}
where $\gamma_{1_0}>0$ and $R^*_0>0$ are nominal values of $\gamma_1$ and $R^*$, respectively. The well-rate control signal is generated by
\begin{equation}
\begin{split}
  Q_c(t)&=\underbrace{B_0^+\left[-k_1\Sabs{\sigma_R(t)}^{\frac{1}{1-l}}+\nu(t)\right]}_{\text{(I) SR-regulation term}},\\
  \dot\nu(t)&=\underbrace{-k_2\Sabs{\sigma_R(t)}^{\frac{1+l}{1-l}}}_{\text{(I) internal SR-regulation term}}
  \underbrace{-k_3\rho\nu(t)}_{\text{(II) saturation-handling term}},
\end{split}
\label{eq:Q}
\end{equation}
where $k_1,k_2>0$, $k_3>1$, $l\in[-1,0]$, and $\Sabs{z}^{\gamma}=|z|^\gamma\operatorname{sign}(z)$. The symbol $(\cdot)^+$ denotes the right pseudoinverse, and
\begin{equation}
  B_0=-\frac{1}{\beta_0|V|}\mathbb{1}^{1\times n},
  \label{eq:B0}
\end{equation}
where $\beta_0>0$ is the nominal compressibility and $\mathbb{1}^{1\times n}$ is the row vector of ones. The saturation-dependent coefficient is
\begin{equation}
  \rho=
  \begin{cases}
    0, & Q_{m_i}<Q_i<Q_{M_i}\ \text{for all }i,\\
    \dfrac{k_2\|B_0^+\|_\infty}{|Q_{M_i}-Q_{m_i}|},
    & \text{if at least one component is saturated}.
  \end{cases}
  \label{eq:rho}
\end{equation}

The first terms (I) in \eqref{eq:Q} are responsible for regulating the seismicity-rate output by driving the tracking error $\sigma(t)$ towards zero. The controller belongs to the super-twisting family of sliding-mode controllers \citep{b:Mathey-Moreno-2024}. The parameter $l$ allows one to interpolate between different behaviors: for $l=-1$ the controller corresponds to the classical (MIMO) super-twisting algorithm, whereas for $l=0$ it degenerates to a linear integral-type regulator.

This structure is widely used because it is robust to modeling errors and bounded disturbances while producing a \emph{continuous} control signal, which is important for field implementation. In the reservoir setting, modeling errors arise from uncertain and heterogeneous hydraulic properties (e.g., $k$, $\eta$, $\beta$) and from simplified SR physics with poorly identified, possibly time-varying parameters (e.g., $R^*$, $\gamma_1$, $\gamma_2$) due to evolving stress conditions and compaction. Disturbances include mismatches between commanded and realized well rates (actuator dynamics, saturation) and noisy SR feedback inferred from discrete catalogs (counting variability, detection incompleteness). Under the assumptions and gain conditions of the nominal analysis, the tracking property is preserved for the admissible class of matched bounded perturbations acting on the regulated channel \citep{b:Gutierrez-Stefanou-2025}. The super-twisting design is therefore attractive because it maintains good tracking under these bounded uncertainties while keeping the well-rate command continuous.

The matrix $B_0^{+}$ plays the role of distributing the SR-regulation action across the wells, using only nominal information about the plant. In other words, it converts a single regulation effort (built from $\sigma$ and $\nu$) into a physically realizable vector of well-rate adjustments. The choice of $B_0$ follows directly from the nominal compressibility $\beta_0$ in \eqref{eq:B0}. The corresponding selection and its rationale are detailed in \cite{b:Gutierrez-Stefanou-2025}.

As already mentioned, each well-rate component is subject to actuator limits, so the commanded $Q_c(t)$ may be saturated. Under saturation, the integral internal state, $\nu$, may be always increasing (or decreasing), leading to sluggish recovery or overshoot once the actuator returns to its admissible range. This is known as ``wind up'' \cite{b:AstromRundqwist1989Windup}. To mitigate this effect, we introduce the additional damping (``leakage'') term (II) in the internal dynamics of $\nu$. This modification is inspired by anti-windup strategies used for super-twisting actuators \citep{b:GOLKANI201852,b:9029793}, and it helps keep $\nu$ bounded and responsive when saturation is active. 

A practical advantage of this controller is that it requires only limited model information: it uses the measured output $y_R(t)$ and a nominal matrix $B_0$, rather than a full high-fidelity model of the reservoir-seismicity system. 

With a suitable choice of gains $k_1$ and $k_2$, the terms (I) of this controller drive the tracking error \eqref{eq:error} to zero (when $l=-1$) or to a small neighborhood of zero (when $l\in(-1,0)$), when there is no saturation (see \cite{b:Gutierrez-Stefanou-2025} for the theoretical details). For the case of saturated fluxes, which is the case of the present work, the added term (II) is ensured to preserve the former results for $k_3>1$ and sufficiently large $\abs{Q_{M_i}-Q_{m_i}}$ (see \cite{b:GOLKANI201852,b:9029793}).

A practical tuning sequence is as follows. First, $k_1$ and $k_2$ are selected inside the admissible super-twisting gain region determined by the nominal input gain and the assumed perturbation bounds. Their common scale is then adjusted to obtain the desired response speed and control effort. Finally, $k_3$ is tuned only after saturation is included: larger values accelerate leakage and recovery from saturation but reduce the accumulated regulation action, thereby changing the production--seismicity compromise. Once stability conditions are satisfied, the gains may also be optimized according to application-specific objectives. Reinforcement learning has, for example, been used to tune robust-controller gains for an underground-reservoir model \citep{b:https://doi.org/10.1002/nag.3923}.

\subsubsection{Digital implementation of the control}

Finally, to reflect the digital (sampled-data) implementation of the controller, the continuous-time command $Q_c(t)$ is applied through a \emph{zero-order hold} (ZOH). Concretely, for a chosen sampling period $\Delta t_c>0$, the sampling instants are defined as $t_k := t_0 + k\Delta t_c$, $k=0,1,2,\dots$ At each sampling instant $t_k$, the controller is evaluated using the current measurement $y_R(t_k)$ to produce the discrete command $Q_c[k] := Q_c(t_k)$. Under ZOH actuation, this command is held constant between updates, i.e., $Q_c(t) = Q_c[k]$ for $t\in[t_k,t_{k+1})$. Therefore, the applied well rates are piecewise constant in time, resulting in a staircase signal, as commonly assumed in digital control implementations (see, e.g., \cite{b:AstromWittenmark1997,b:ChenFrancis1995} for background on sampled-data and ZOH actuation).

Sampling and ZOH inevitably introduce a small tracking and implementation error compared with the ideal continuous-time controller. In practice, the main qualitative properties of super-twisting-type controllers are typically retained provided that the sampling is \emph{fast enough} relative to the closed-loop dynamics, i.e., $\Delta t_c$ is sufficiently small compared with the dominant time scales of the plant and the intended controller convergence time. As $\Delta t_c$ increases, the controller reacts with a delay and the tracking error increases \citep{b:8450003}. In the present application, substantial seismicity mitigation is maintained across the investigated sampling periods, from one day to seven months, as shown in Fig.~\ref{fig:Dtc}.

The controller is formulated in continuous time in \eqref{eq:Q}. Although a sampled-data implementation can be obtained through a ZOH, \emph{proper} discrete-time realizations of the super-twisting algorithm may also be adopted \citep{b:8896941,b:10449369,b:SEEBER2025112027}. Extending such discretizations to the present setting is a natural direction, but it is beyond the scope of this paper.

\subsection{Closed-loop numerical implementation}
\label{sec:numerics}

The simulations were implemented in Python. Finite-element matrices were assembled with \texttt{scikit-fem} \citep{GustafssonMcBain2020}, and the semi-discrete pressure--SR dynamics were integrated with SciPy's \texttt{solve\_ivp} using the BDF method \citep{VirtanenEtAl2020}. At each control interval, the closed loop is executed as follows:
\begin{enumerate}
  \item hold the previously computed total well-rate command
  \eqref{eq:control+static} over $[t_k,t_{k+1})$,

  \item integrate the pressure-diffusion model
  \eqref{eq:diff}--\eqref{eq:B} and the SR dynamics
  \eqref{eq:SR1}--\eqref{eq:SR} over the interval,

  \item sample the event count and times from
  \eqref{eq:event_rate}--\eqref{eq:cum_intensity}, the event locations
  from \eqref{eq:conditional_density}, and the magnitudes from \eqref{eq:GR},

  \item count the events in the measurement window and update the raw
  and filtered SR estimates according to
  \eqref{eq:raw}--\eqref{eq:ema_disc},

  \item compute the tracking error \eqref{eq:error}, update the controller
  state, and calculate the new well-rate command using
  \eqref{eq:Q}--\eqref{eq:rho}. For Scenario~2, use instead the production
  constraint \eqref{eq:restriction} and the constrained controller
  \eqref{eq:Q2}, and

  \item impose the componentwise rate limits according to
  \eqref{eq:sat} and hold the resulting command over the next interval.
\end{enumerate}
The reservoir assembly, event generator, converter, constrained controller, closed-loop update logic and related algorithms were developed by the authors. Reproducibility scripts will be released through the ERC-INJECT GitHub upon publication.

\section{Results}
\label{sec:Results}

Two control scenarios are investigated, starting from the onset of recorded induced seismicity in December, 1991. 

\textbf{Scenario 1.} All wells are assumed to operate in extraction only. The controller \eqref{eq:Q} is used to regulate the reservoir's seismicity while, as far as possible, maintaining the historical total gas-extraction demand shown in Fig.~\ref{fig:extraction}.

\textbf{Scenario 2.} A hypothetical subset of wells is for injection, while the remaining wells exactly track the sampled historical gas-extraction demand shown in Fig.~\ref{fig:extraction}. The controller is also applied to the wells to regulate the reservoir's seismicity, subject to the same operational constraints. This partition is a conceptual control-allocation assumption and not an existing Groningen configuration.

\subsection{Scenario 1: gas extraction}

To enforce the production-only (extraction) scenario, we set the saturation bounds in \eqref{eq:sat} to $Q_{M_i}=0$~[10$^{6}$ m$^{3}$/month] and $Q_{m_i}=-511.35$~[10$^{6}$ m$^{3}$/month] (for all $i=1,...,29$), so that each well is constrained to non-positive flux (extraction). The total well-rate input in \eqref{eq:control+static} is then implemented as the sum of the historical (static) extraction profile $Q_s(t)$ and the feedback correction $Q_c(t)$ defined in \eqref{eq:Q}. In this configuration, $Q_s(t)$ enforces the target production demand from Fig.~\ref{fig:extraction}, while $Q_c(t)$ redistributes the well rates (within the saturation limits) to reduce induced seismicity.

The tracking error \eqref{eq:error} is computed using $\gamma_{1_0}=2 \gamma_{1}$ and
$R^*_0=\frac{1}{\abs{V}}\int_V R^*(x)\,dV = 4.11\times 10^{-5}\ \text{[events/(km$^{3}\cdot$year)]}$. The desired SR reference is selected as \(r_R(t)=R_0^*\). The nominal matrix $B_0$ is selected as in \eqref{eq:B0}, with $\beta_0=0.8\,\beta$. Notice that $\gamma_{1_0}$ and $\beta_0$ were selected in this way to show the robustness of the method against parameter uncertainties. The controller parameters are chosen as $l=-1$ (super-twisting), $k_1=6.7\times 10^{-5}$, $k_2=2.2\times 10^{-9}$ and $k_3=100$. The converter \eqref{eq:ema_disc} was applied with $\tau=1$ [month]. The control signal $Q_c$ was digitally implemented with a sampling period of $\Delta t_c=1$ [month], i.e., the control is updated every month and it is kept constant during this period ($Q_s$ is not).

\begin{figure*}[pos=tbp]
  \centering
  \includegraphics[width=11.cm,keepaspectratio]{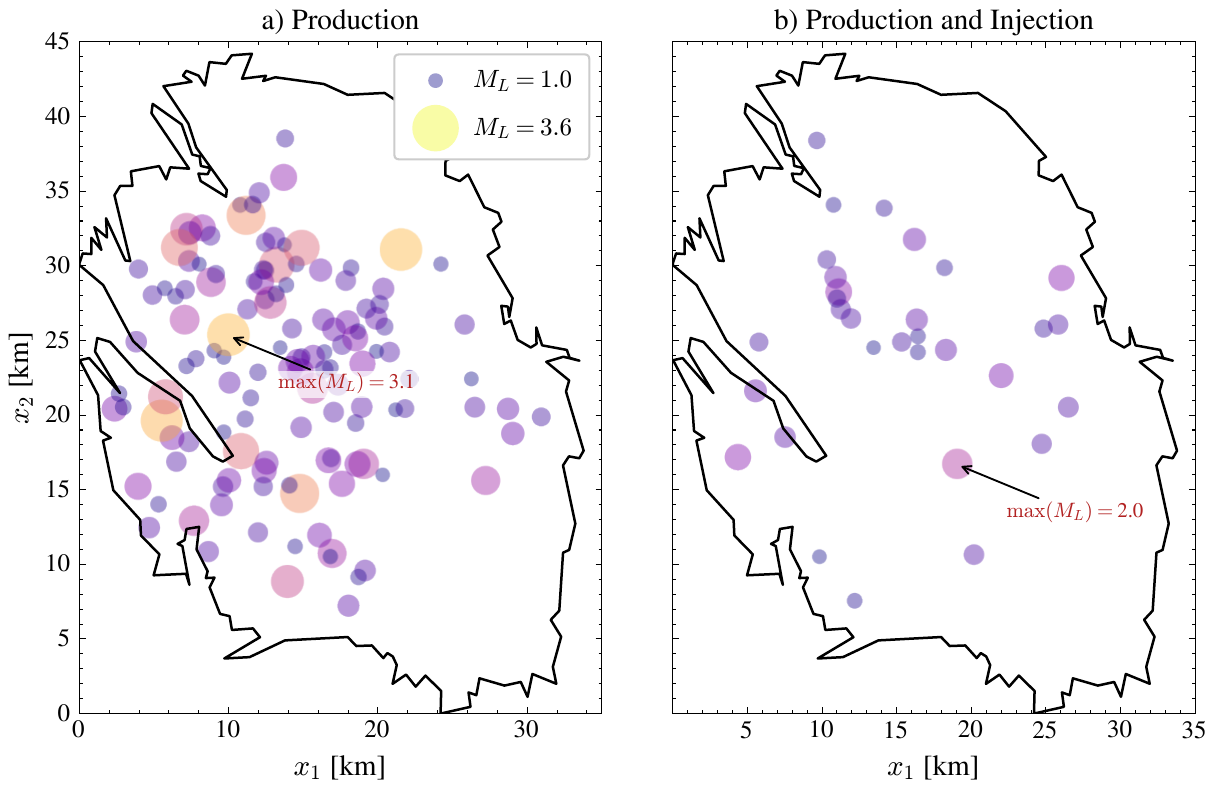}
  \caption{Example controlled catalogs at the end of the simulation. (a) Extraction-only operation, where feedback mitigation may curtail the historical production profile. (b) Conceptual production--injection operation, where production follows the prescribed demand and separate injection channels regulate SR. Marker size and color represent local magnitude.}
  \label{fig:catalog2}
\end{figure*}

\begin{figure}[pos=htbp]
  \centering
  \includegraphics[width=8.5cm,keepaspectratio]{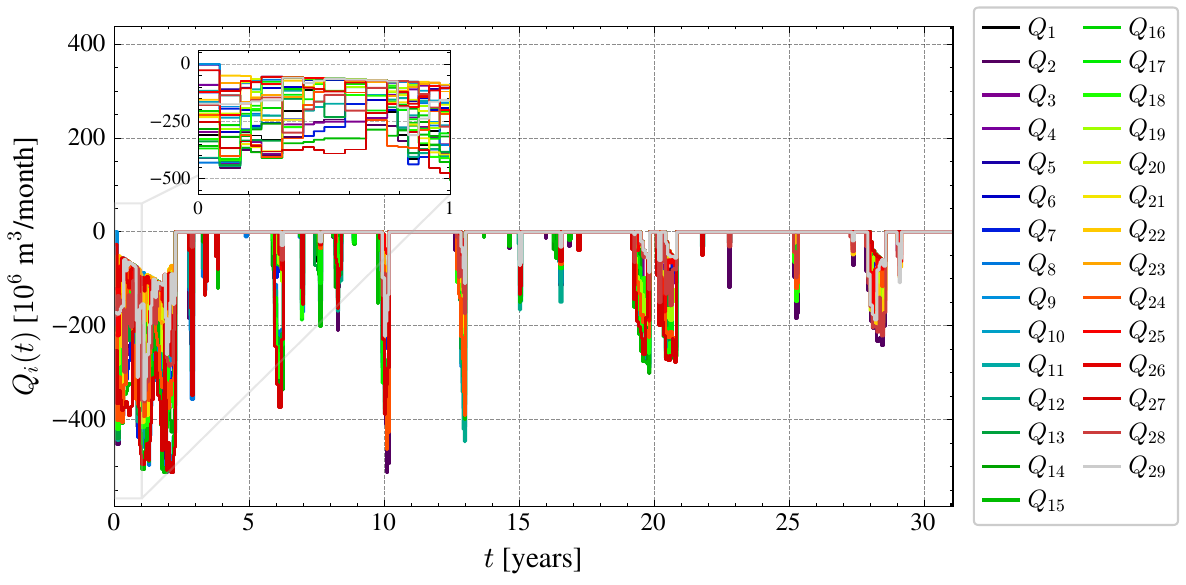}
  \caption{Individual bounded well rates in the extraction-only scenario. The inset highlights the staircase profile produced by the monthly ZOH update.}
  \label{fig:Q1}
\end{figure}

\begin{figure}[pos=htbp]
  \centering
  \includegraphics[width=8.0cm,keepaspectratio]{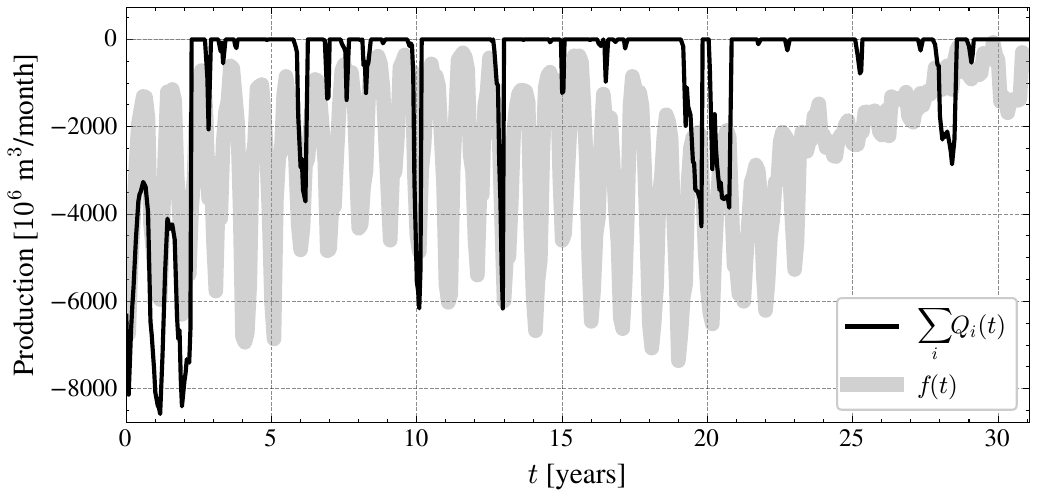}
  \caption{Total extraction under feedback and the historical no-control reference for $k_3=100$ and $\Delta t_c=1$~month. SR mitigation is obtained by redistributing and, when required, reducing extraction.}
  \label{fig:D1}
\end{figure}

Figure~\ref{fig:catalog2}(a) shows a controlled catalog. Relative to the no-control realizations in Fig.~\ref{fig:catalog}, substantially fewer events are generated. Figure~\ref{fig:Q1} shows the bounded piecewise-constant well rates, and Fig.~\ref{fig:D1} compares their total with the historical reference. The controller initially follows the prescribed extraction more closely, but subsequently reduces total extraction when the catalog-based SR estimate rises. Strong mitigation therefore requires deviation from the historical demand in this extraction-only configuration.

\begin{figure}[pos=htbp]
    \centering
    (a)\includegraphics[width=8.0cm,keepaspectratio]{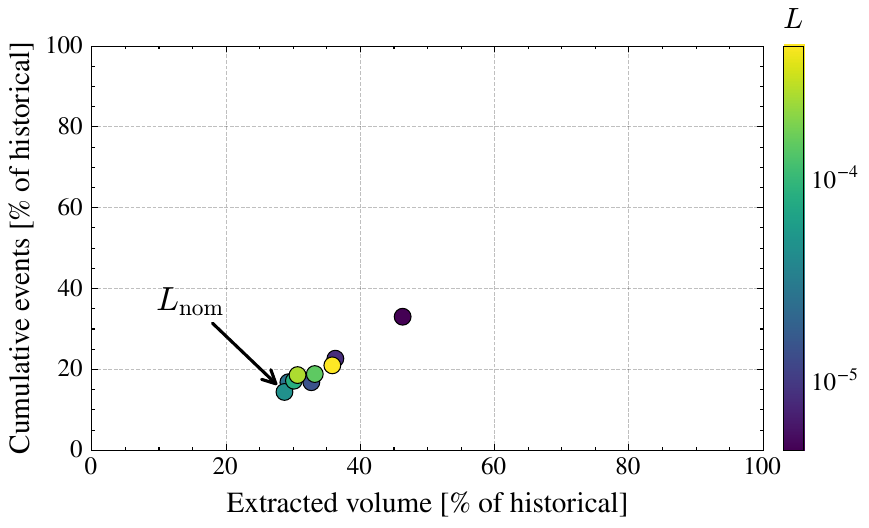}
    (b)\includegraphics[width=8.0cm,keepaspectratio]{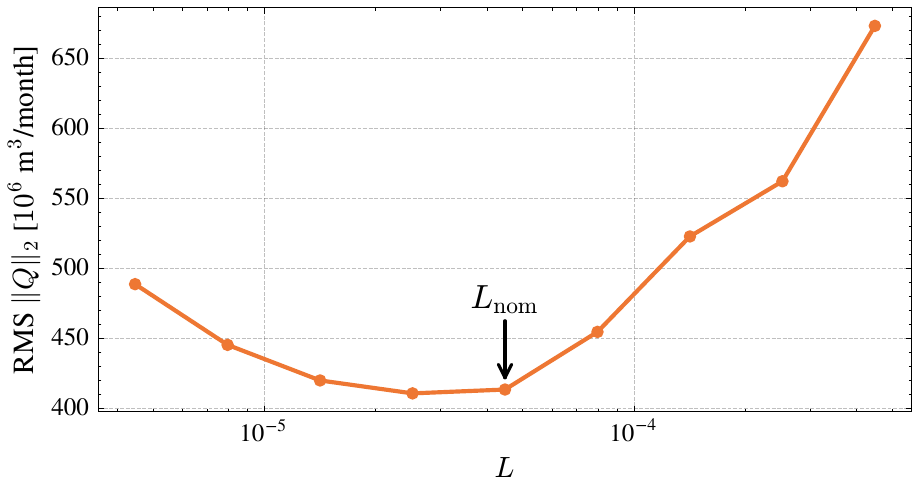}
    \caption{Sensitivity to the super-twisting gain scaling $L$: (a) final cumulative event count versus extracted gas volume, with colors indicating $L$; (b) RMS magnitude of the total applied well-rate vector versus $L$. The leakage gain $k_3=100$ and sampling period $\Delta t_c=1$ month are fixed.}
    \label{fig:L}
\end{figure}

To assess the sensitivity to the super-twisting gains $k_1$ and $k_2$, we consider a homogeneous scaling using a single parameter $L>0$, following \cite{b:Mathey-Moreno-2024}, $k_1(L)=\frac{L}{L_{\mathrm{nom}}}k_{1,\mathrm{nom}}$, $k_2(L)=\left(\frac{L}{L_{\mathrm{nom}}}\right)^2 k_{2,\mathrm{nom}}$, where $k_{1,\mathrm{nom}}=6.7\times 10^{-5}$, $k_{2,\mathrm{nom}}=2.2\times 10^{-9}$, $L_{\mathrm{nom}}=4.47\times10^{-5}$ correspond to the nominal gains used in the previous simulations. We consider 9 logarithmically spaced values of $L\in[4.47\times10^{-6},4.47\times10^{-4}]$, keeping $k_3=100$ and the sampling period fixed. Figure~\ref{fig:L}(a) shows that substantial seismicity mitigation is maintained throughout the investigated range of $L$, although the resulting event counts and extracted volumes vary nonmonotonically with the gain. Figure~\ref{fig:L}(b) reveals a nonmonotonic dependence of the RMS applied well-rate vector on $L$, with a minimum close to the nominal configuration and a pronounced increase for larger values. These results indicate that increasing the super-twisting gains does not necessarily improve the production--seismicity compromise and may increase the required actuation effort.

\begin{figure}[pos=htbp]
  \centering
  \includegraphics[width=8.0cm,keepaspectratio]{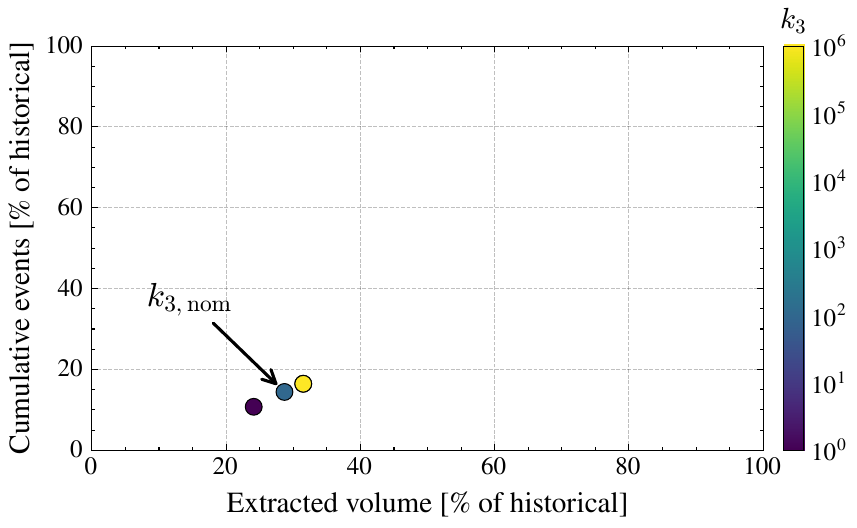}
  \caption{Final cumulative event count versus extracted gas volume for different leakage gains $k_3$. Each point represents one closed-loop simulation with identical physical and sampling parameters. Points $k_3=10^4$ and $k_3=10^6$ overlap.}
  \label{fig:k3}
\end{figure}

The leakage gain $k_3$ governs the compromise between preserving accumulated regulation action and recovering from saturation. Figure~\ref{fig:k3} shows that increasing $k_3$ produces a moderate increase in the cumulative extracted volume, accompanied by a comparatively small increase in seismicity. The response becomes less sensitive to further increases in $k_3$, with nearly coincident results for the largest gains considered. Therefore, $k_3$ provides an additional tuning parameter, although its influence is less pronounced than that of the super-twisting gain scaling $L$ over the investigated ranges.

\begin{figure*}[pos=htbp]
    \centering
    (a)\includegraphics[width=8.0cm,keepaspectratio]{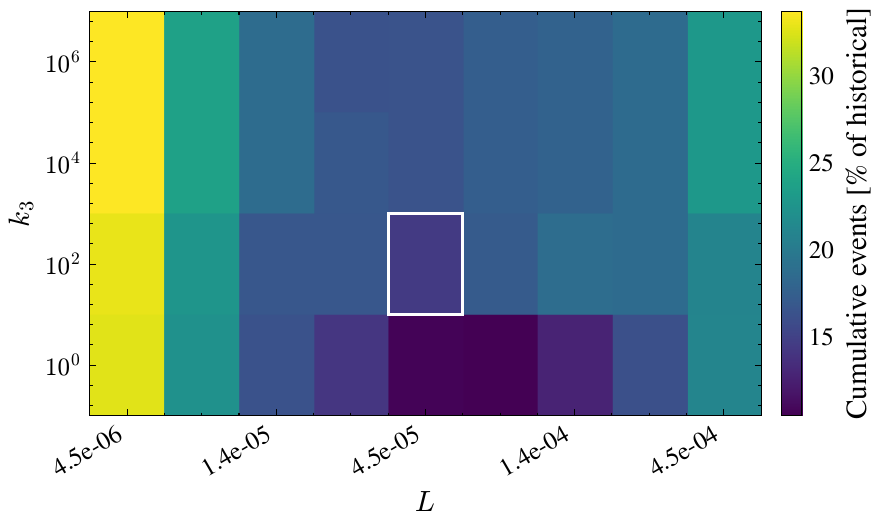}
    (b)\includegraphics[width=8.0cm,keepaspectratio]{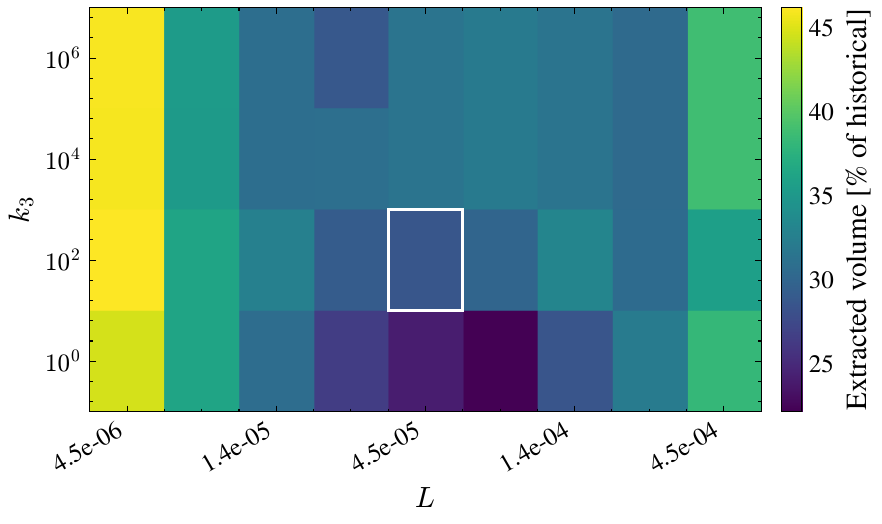}
    (c)\includegraphics[width=8.0cm,keepaspectratio]{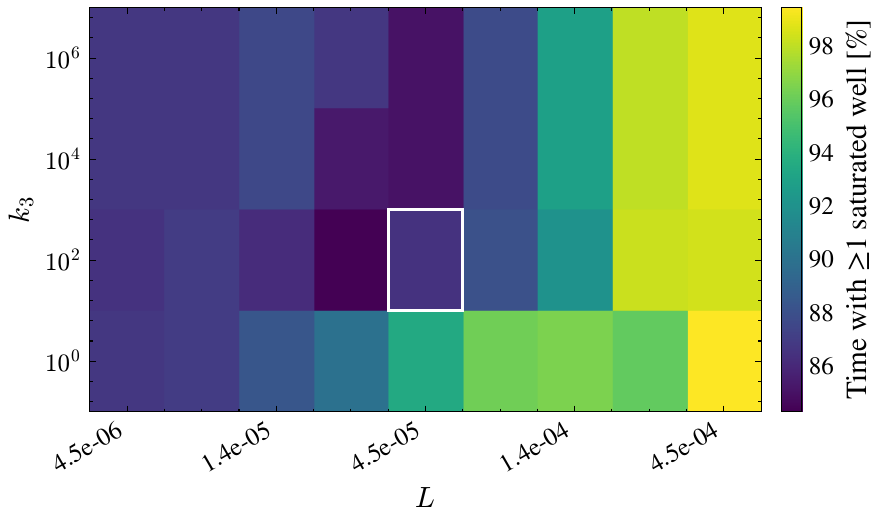}
    (d)\includegraphics[width=8.0cm,keepaspectratio]{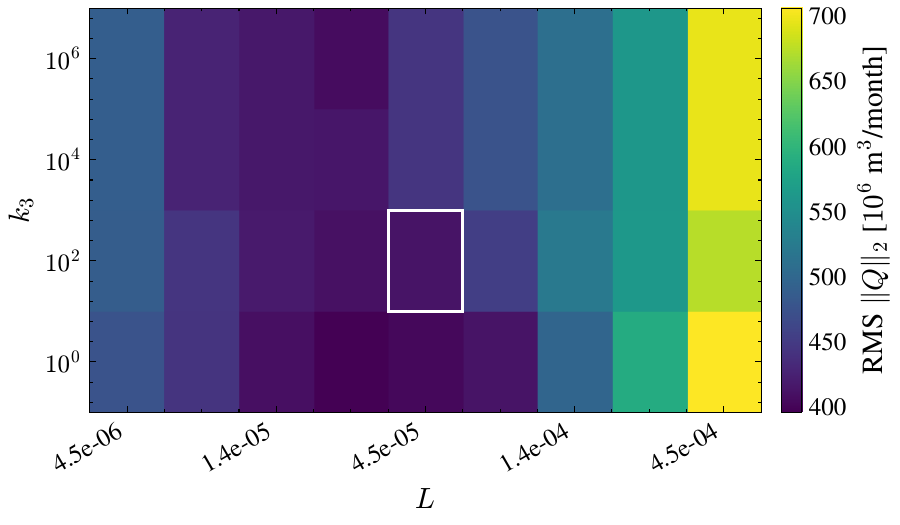}
    \caption{Joint sensitivity to the super-twisting gain scaling $L$ and the integral leakage gain $k_3$: (a) final cumulative event count relative to the historical catalog; (b) cumulative extracted gas volume relative to historical production; (c) fraction of time with actuator saturation; and (d) RMS magnitude of the total applied well-rate vector. The sampling period is fixed at $\Delta t_c=1$ month. The white rectangle corresponds to the nominal values used in the rest of the simulations.}
    \label{fig:heat}
\end{figure*}

For completeness, the combined influence of the super-twisting gain scaling $L$ and the integral leakage gain $k_3$ is investigated through a sensitivity analysis. Nine logarithmically spaced values of $L\in[4.47\times10^{-6},4.47\times10^{-4}]$ and four values of $k_3\in\{10^0,10^2,10^4,10^6\}$ are considered, resulting in 36 configurations. The same stochastic seed and a sampling period of $\Delta t_c=1$ month are used throughout. Figure~\ref{fig:heat} illustrates the resulting cumulative seismicity, extracted gas volume, actuator saturation frequency, and control effort, providing an overview of the trade-offs associated with the selection of these gains. Figure~\ref{fig:heat} further shows that the influence of \(L\) is more pronounced than that of \(k_3\) over the investigated ranges, particularly regarding extracted volume and RMS actuation effort. Actuator saturation remains frequent across the configurations, with at least one well reaching an operational bound during most of the simulation. Nevertheless, substantial seismicity mitigation is maintained, demonstrating the importance of explicitly accounting for well-rate constraints.

\begin{figure}[pos=htbp]
  \centering
  \includegraphics[width=8.0cm,keepaspectratio]{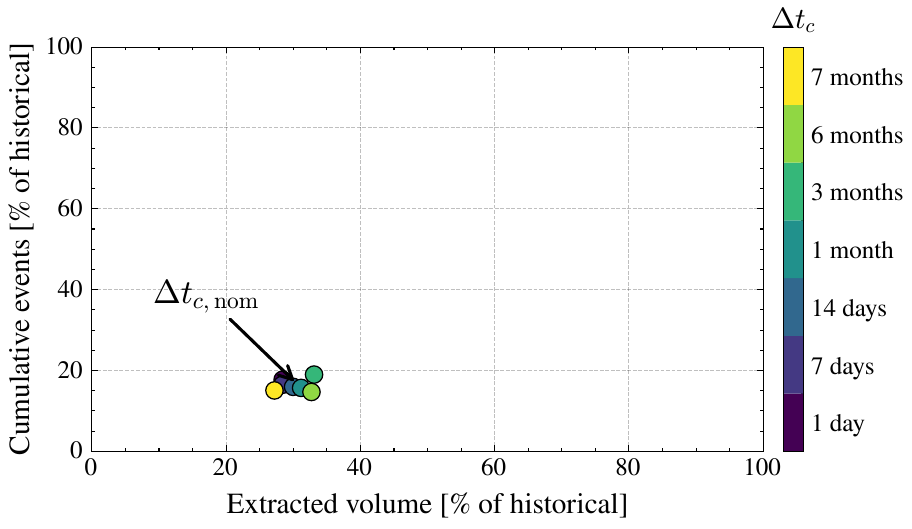}
  \caption{Effect of the ZOH update period on final cumulative events and extracted volume for fixed nominal gains. The results remain within a relatively narrow range over the investigated sampling periods.}
  \label{fig:Dtc}
\end{figure}

Figure~\ref{fig:Dtc} assesses the update period for fixed nominal gains. The results show relatively limited variations in cumulative seismicity and extracted volume across the investigated sampling periods, from one day to seven months. This suggests that, within the considered configuration, the closed-loop performance is relatively insensitive to the sampling period over this range.

\begin{figure}[pos=htbp]
  \centering
  \includegraphics[width=8.0cm,keepaspectratio]{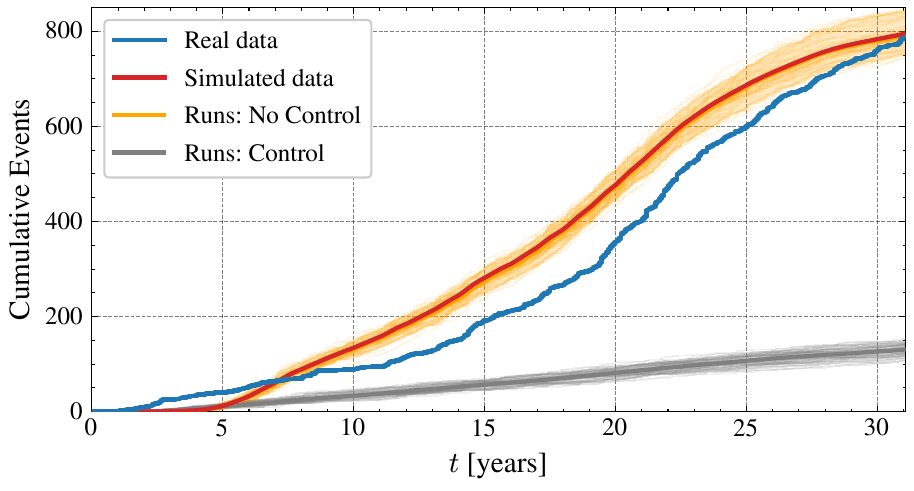}
  \caption{Cumulative event counts for 50 independent realizations of each case: no control and extraction-only feedback control. The ensemble means and shaded pointwise 5th--95th percentile prediction bands are shown together with the observed cumulative event count and the deterministic no-control reference.}
  \label{fig:runs}
\end{figure}

Stochastic variability is assessed using 50 independent realizations for the no-control and extraction-only feedback cases. Figure~\ref{fig:runs} shows the ensemble means and pointwise 5th--95th percentile bands. Table~\ref{tab:ensemble} summarizes the final event count, maximum magnitude, and spatial-density correlation. To quantify the spatial variability across stochastic realizations, we also show the normalized spatial RMS uncertainty 
$    U_s = \left(
    \frac{\int_V s_d(x)^2\,\mathrm{d}V}
    {\int_V \bar d(x)^2\,\mathrm{d}V}
    \right)^{1/2}$, where $\bar d(x)$ and $s_d(x)$ denote the ensemble-mean spatial density and its pointwise standard deviation, respectively. The controlled deterministic reference is averaged across the run-dependent deterministic trajectories because each catalog changes the feedback signal and hence the closed-loop SR field.

\begin{table}[pos=htbp]
\centering
\caption{Ensemble statistics from 50 independent stochastic catalogs. Intervals represent the 5th to 95th percentile of the simulated results.}
\label{tab:ensemble}
\small
\resizebox{\columnwidth}{!}{%
\begin{tabular}{lcc}
\hline
Statistic & No Control & Scenario 1 \\
\hline
Deterministic events
    & $795.11$ & $130.71$ \\
Final events, mean$\pm$SD
    & $791.76 \pm 27.36$ & $131.08 \pm 11.16$ \\
Relative difference
    & $-0.42\%$ & $0.28\%$ \\
Final-event simulated interval
    & $[752.35, 842.10]$ & $[111.15, 147.20]$ \\
Maximum magnitude, mean$\pm$SD
    & $3.47 \pm 0.11$ & $3.16 \pm 0.28$ \\
Magnitude simulated interval
    & $[3.25, 3.60]$ & $[2.71, 3.55]$ \\
Density correlation
    & $0.9776$ & $0.9665$ \\
Normalized spatial uncertainty
	& $11.12\%$ & $22.48\%$ \\
\hline
\end{tabular}%
}
\end{table}

The reduced maximum magnitude under feedback does not indicate a change in the prescribed magnitude law: both cases use the same truncated Gutenberg--Richter distribution. It follows statistically from sampling fewer events. 

The results show that feedback substantially reduces the mean number of seismic events and their variability, together with a reduction in the sampled maximum magnitudes. Both scenarios exhibit high spatial-density correlations and moderate normalized spatial uncertainties, although the latter is higher under feedback, consistent with the smaller number of generated events. Overall, these results support the effectiveness of the proposed catalog-based feedback strategy and its performance under stochastic seismicity within the considered numerical model.

\subsection{Scenario 2: combined production and injection}

Scenario~2 is a conceptual control-allocation study. The model locations are divided into extraction-only and injection-only groups distributed across the field, as indicated in Fig.~\ref{fig:Groningen}, to provide spatial coverage of both actuation types. The assignment is used only in the simulations and does not imply that the corresponding Groningen production locations are injection-ready. The saturation bounds in \eqref{eq:sat} are $Q_{M_i}=511.35$~[10$^{6}$ m$^{3}$/month] and $Q_{m_i}=0.0$~[10$^{6}$ m$^{3}$/month] for the injection wells, and $Q_{M_i}=0.0$~[10$^{6}$ m$^{3}$/month], $Q_{m_i}=-511.35$~[10$^{6}$ m$^{3}$/month] for the extraction (production) wells. The total well-rate input in \eqref{eq:control+static} is implemented solely through the feedback term $Q_c(t)$.

To ensure that the overall gas production matches the historical extraction demand, $f(t)$, an additional \emph{production constraint} is imposed on the control input following \citep{b:Gutierrez-Stefanou-2024,b:Gutierrez-Stefanou-2025}. Specifically, a weighted sum of the well rates is required to match a prescribed time-varying demand,
\begin{equation}
    W\,Q_c(t) = f(t),
    \label{eq:restriction}
\end{equation}
where $W \in \mathbb{R}^{1 \times 29}$ is a full-rank row vector encoding which wells contribute to gas production. In the present implementation, $W$ has entries equal to one for extraction-only wells and zero for injection-only wells. Therefore, \eqref{eq:restriction} enforces that the total extraction from the production wells equals the target profile $f(t)$.

Under this constraint, the control law is redefined as
\begin{equation} 
\begin{split} 
Q_c(t) &= \underbrace{\overline{W}\left(B_0 \overline{W}\right)^{+}\Big[-k_1 \Sabs{\sigma(t)}^{\frac{1}{1-l}} + \nu(t)\Big]}_{\text{(I) SR-regulation term}} \\ 
&\quad \underbrace{+W^T\left(W W^T\right)^{-1}f(t)}_{\text{(III) Constraint-tracking term}},\\ 
\dot{\nu}(t) &= \underbrace{-k_2 \Sabs{\sigma(t)}^{\frac{1+l}{1-l}}}_{\text{(I) Internal SR-regulation term}}
\;\;\underbrace{-k_3\,\rho\,\nu}_{\text{(II) Saturation handling term}}. 
\end{split} 
\label{eq:Q2} 
\end{equation}
Here, $\overline{W}\in\mathbb{R}^{29\times 28}$ is a basis for the null space of $W$, so that $W\overline{W}=0$. The third term (III) in \eqref{eq:Q2} enforces the operational requirement that the total injection/production satisfy the prescribed constraint, or production target, $f(t)$. This term distributes $f(t)$ among the wells according to the weights in $W$, without interfering with the SR-regulation action, which is handled by the first terms. In practice, this ensures that the controller respects the production plan while terms (I)--(II) regulate seismicity.

Figure~\ref{fig:catalog2} b) shows the synthetic catalog obtained for Scenario~2, with the same control parameters as Scenario~1. Compared to Scenario~1, the reduction in induced seismicity is markedly stronger. This improvement stems from the additional operational flexibility provided by compensating injection: in this setting, injection acts as a stabilizing mechanism for the SR dynamics, enabling the controller to counterbalance pressure-driven stressing associated with production more effectively.

\begin{figure}[pos=htbp]
  \centering
  \includegraphics[width=8.5cm,keepaspectratio]{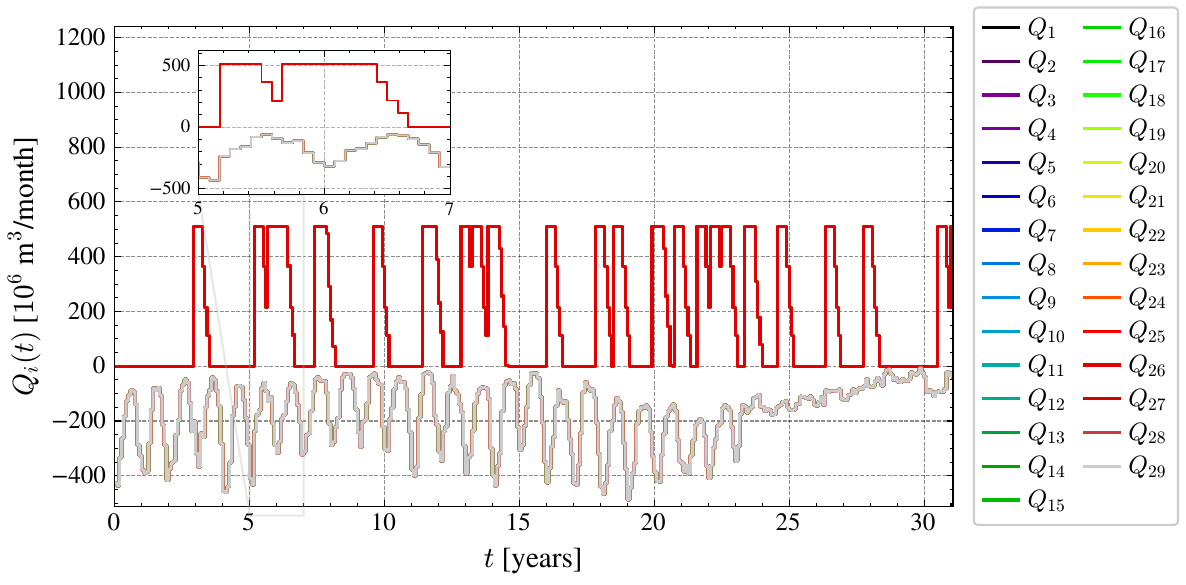}
  \caption{Individual well rates in the conceptual production--injection scenario. Production-only locations satisfy the prescribed demand, whereas injection-only locations provide independent SR-regulation authority. The inset shows the ZOH implementation.}
  \label{fig:Q2}
\end{figure}

The injection and extraction well fluxes, $Q(t)$, are depicted in Figure~\ref{fig:Q2}, where approximately half of the wells inject fluid to mitigate induced seismicity. The identical fluxes observed within each group of wells (production and injection) result from considering a single reservoir-average SR output, and hence a single tracking error, together with the uniform control allocation defined by $B_0$ and the production constraint encoded by $W$. This differs from Scenario 1, where the feedback correction is superimposed on the individual historical extraction profiles, which differ among wells. Despite this relatively simple allocation strategy, substantial seismicity mitigation is achieved while preserving the prescribed production demand. Other configurations involving multiple regional SR outputs could be explored through an optimization analysis to determine the most suitable regional partition and well allocation according to selected cost functions. Such an analysis is, however, beyond the scope of the present work.

\begin{figure}[pos=htbp]
  \centering
  \includegraphics[width=8.0cm,keepaspectratio]{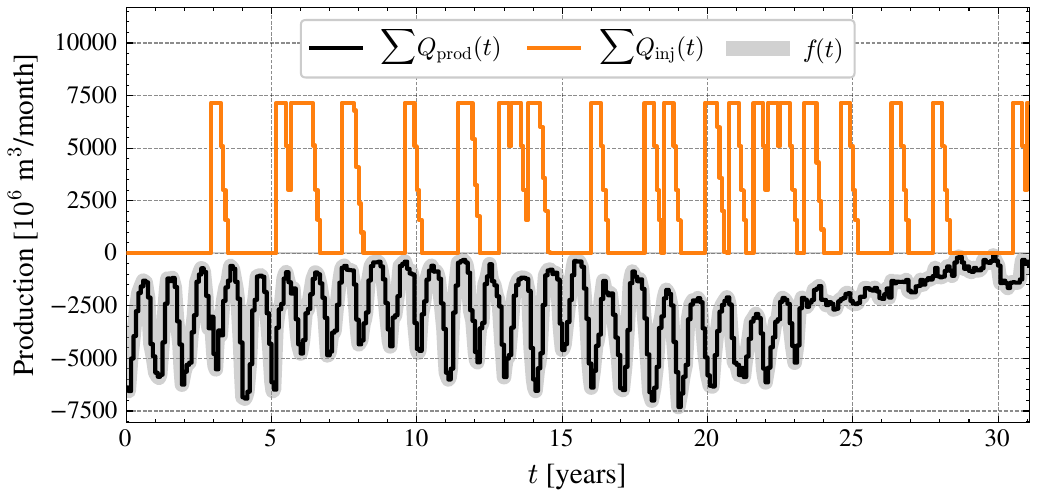}
  \caption{Scenario~2 totals: extraction from the production subset, injection from the injection subset, and historical production demand $f(t)$. Constraint~\eqref{eq:restriction} enforces exact tracking of the sampled production demand, while the null-space action is used for SR regulation.}
  \label{fig:D2}
\end{figure}

Figure~\ref{fig:D2} shows the total flux extracted by the production wells, the total flux injected by the injection-only wells, and the reference production target $f(t)$. By construction of the constraint \eqref{eq:restriction}, the controller achieves exact tracking of the sampled historical gas-production demand using only the production-well subset, while the injection-well subset is used to regulate seismicity and reduce the number of events.

\section{Discussion}
\label{sec:Discussion}

The historical production profile without feedback provides the direct baseline for the present model. This record seems to include periods of seismicity-related production restrictions, notably under the Groningen \emph{Meet- en regelprotocol} introduced in 2017 \citep{SodM2017Protocol,SodM2018Zeerijp}. Thus, the reference already represents historical operation of gas production rather than an unrestricted extraction scenario (see also Fig.~\ref{fig:extraction}). Relative to this baseline, the presented feedback substantially reduces the ensemble-mean event count while making the associated gas extraction explicit.

Operationally oriented Groningen studies have examined production measures, optimized production distributions, forecasting, and decision support \citep{MuntendamBosEtAl2017,SodM2017Optimisation,BourneOatesVanElk2018,b:SMITH2022117697}. The official \emph{Meet- en regelprotocol} used monitored seismic indicators and prescribed signaling and intervention actions \citep{SodM2017Protocol,SodM2018Zeerijp}, while \cite{Jansen2018OEFS} proposed operational earthquake forecasting as a testbed for adaptive traffic-light systems and optimized production or reinjection schedules. Such methods are complementary but structurally different: threshold protocols trigger predefined interventions, whereas \eqref{eq:Q} updates the well rates from the measured SR error while enforcing rate constraints. A direct quantitative comparison would additionally require specifying the thresholds, decision horizons, restart rules, and individual well-rate allocations of a particular intervention strategy and, therefore, is omitted.

The extraction-only trade-off follows directly from the no-flow boundary condition. Indeed, integrating \eqref{eq:diff} over $V$ and using the boundary conditions yields $\beta\int_V u_t(x,t)\,\mathrm{d}x=\sum_{i=1}^{n}Q_i(t)$, showing that a prescribed nonzero net extraction necessarily produces a nonzero domain-average pressure rate. Since \eqref{eq:SR} uses this pressure rate as a driving input, strict historical production tracking and strong SR suppression are competing objectives. Scenario~1 therefore represents an achievable compromise: feedback reallocates and curtails extraction as required, but cannot preserve both objectives exactly. A drained or leaky boundary would add a boundary-flux contribution and could change the quantitative trade-off. The feedback architecture itself is not restricted to the no-flow boundary condition case and can be implemented with zero pressure condition as well \citep{b:Gutierrez-Stefanou-2025}.

The stochastic experiments show that the controller can close the loop using event counts rather than direct access to the deterministic SR state. Filtering, sampling, saturation, and gain selection nevertheless affect performance. A longer filter horizon reduces catalog noise but increases delay, a longer update period delays corrective action, and the leakage gain changes the recovery from saturation and the retained integral action. The sensitivity analysis reveals that the production--seismicity compromise depends on the controller gains in a nonmonotonic manner, while relatively small variations are observed across the investigated control update periods. This compromise is an emergent property of the specific setup and boundary conditions rather than an encoded characteristic of the controller.

Scenario~2 demonstrates the additional control performance supplied by independent injection channels, not the field readiness of nitrogen injection in Groningen. Injection is not intrinsically safe or uniformly stabilizing. Pressure buildup and poroelastic stress transfer can promote fracture propagation or reactivate nearby faults, as illustrated in a different geological setting by \cite{BaoEtAl2025}. Any practical study would require site-specific limits on injection rate, bottom-hole pressure, pressure-rate variation, cumulative injected volume, and injectivity, fault-aware placement and exclusion distances, local pressure and seismic monitoring, automatic safety interlocks, and assessment of well integrity, nitrogen supply and breakthrough, compression and transport infrastructure, and regulatory acceptance. Existing Groningen feasibility studies provide relevant context \citep{NAM2016GPM,Fugro2022KEM24a,OsingaEtAl2026}, but the present rate-saturation model does not intend to establish injection safety or feasibility at a specific location.

A further limitation is the assumed sign/direction of the local control effectiveness: extraction is modeled as increasing SR and injection as decreasing it. In a faulted reservoir, this sign may change with location, stress state, fault geometry, and operating history. If the true direction is opposite to the assumed one, nominal feedback can reinforce the error and destabilize the system. A modular sign-compensation layer for uncertain time-varying input directions has recently been developed \citep{GutierrezOribioStefanou2026Sign}. Integrating this idea with the present stochastic catalog-based architecture is a subject of ongoing work. Such a layer addresses uncertainty in the control direction but does not replace site-specific geomechanical modeling and safety constraints.

Finally, the high spatial calibration correlation should not be interpreted as independent validation because the observed density enters the parameter fit. The reduced-order model omits explicit faults, compaction, poroelastic coupling, three-dimensional layering, and local stress concentrations. The robust-control claim is likewise confined to the bounded matched perturbation class used in the controller analysis and does not compensate for arbitrary omitted physics. The results should therefore be interpreted as evidence for the feasibility and structure of catalog-based constrained feedback, rather than as a field-ready operating prescription.

Overall, the proposed framework could serve as a basis for seismicity-informed operational decision support, allowing well-rate commands to be adapted to observed seismic activity while accounting for production requirements and operational constraints.

\section{Conclusions}
\label{sec:Conclusions}

This paper presented a control-oriented methodology for operating the Groningen gas field while explicitly accounting for induced-seismicity mitigation. The objective was to minimize induced seismicity while preserving operational requirements, such as a prescribed gas-production profile.

To this end, the proposed methodology combines a cascade model coupling pore-pressure diffusion with SR dynamics, a stochastic event-generation step that produces a synthetic earthquake catalog (event times, locations, and magnitudes) consistent with the modeled SR field, and, more importantly, a feedback-control module that estimates the regional SR from the catalog and computes well-rate commands. The proposed controller explicitly accounts for well-flux limitations and digital implementation by applying the control in a discrete-time setting.

The calibrated model reproduces the overall spatial seismicity distribution and final cumulative event count of the Groningen observations, with a cumulative-count NRMSE of 0.090, a final-count relative error of 0.71\%, and a spatial calibration correlation of 0.9999. Across 50 independent stochastic realizations, extraction-only feedback reduced the mean final event count from 791.76 to 131.08, corresponding to an 83.4\% reduction relative to the historical no-control baseline, at the expense of reduced gas extraction. The sensitivity analysis revealed a nonmonotonic dependence of seismicity and control effort on the super-twisting gain scaling $L$, whereas variations in the leakage gain $k_3$ produced comparatively moderate changes. Furthermore, the results exhibited limited sensitivity to the control update period over the investigated range of one day to seven months. These findings demonstrate substantial seismicity mitigation under stochastic feedback, actuator saturation, and sampled-data implementation, while highlighting the importance of controller tuning and the associated production--seismicity compromise.

The numerical results demonstrate SR regulation under the modeled production constraints, well-rate saturation, and sampled-data actuation. Independent production and injection channels allowed exact tracking of the sampled production demand and stronger SR reduction in the conceptual second scenario. This result identifies the control performance provided by compensating injection, but does not establish its operational safety or feasibility in Groningen. Fault-aware pressure constraints, infrastructure assessment, monitoring, and regulatory review would be indispensable.

The main limitations of the present work arise from the reduced-order pressure--stress closure, the absence of explicit faults and three-dimensional geomechanics, and the assumed sign of the control effectiveness. Future work can integrate higher-fidelity reservoir models, further discrete-time controller realizations, and sign-compensation mechanisms while preserving the catalog-based constrained feedback structure.

\section*{Disclaimer}

This study is conducted for academic purposes only. The approaches described are not intended for direct field application or operational decision-making without site-specific validation, calibration, and expert review under applicable safety and regulatory frameworks. Any use beyond this academic context remains the responsibility of the implementing parties.

\printcredits

\section*{Acknowledgments} 

Funded by the European Union. Views and opinions expressed are however those of the author(s) only and do not necessarily reflect those of the European Union or the European Research Council Executive Agency.  Neither the European Union nor the granting authority can be held responsible for them. This work is supported by the ERC grant INJECT, no. 101087771, doi: 10.3030/101087771. Both authors sincerely thank Dr.\ Mateo Acosta, Prof.\ J.-P.\ Avouac, Dr.\ Stephen Bourne, and Dr.\ Jan van Elk for fruitful discussions regarding the Groningen reservoir and related data.

\section*{Declaration of generative AI and AI-assisted technologies}

During the preparation of this work, the author(s) used ChatGPT for language editing. The author(s) reviewed and edited the output as needed and take full responsibility for the content of the published article.

\bibliographystyle{cas-model2-names}

\bibliography{Bibliografias}



\end{document}